# Global Daily CO$_2$ emissions for the year 2020

Zhu Liu[1]*[†], Zhu Deng[1]*[†], Philippe Ciais[2], Jianguang Tan[1], Biqing Zhu[1], Steven J. Davis[3], Robbie Andrew[4], Olivier Boucher[5], Simon Ben Arous[6], Pep Canadell[7], Xinyu Dou[1], Pierre Friedlingstein[8], Pierre Gentine[9], Rui Guo[1], Chaopeng Hong[3], Robert B. Jackson[10], Daniel M. Kammen[11], Piyu Ke[1], Corinne Le Quere[12], Crippa Monica[13], Greet Janssens-Maenhout[13], Glen Peters[4], Katsumasa Tanaka[14], Yilong Wang[15], Bo Zheng[16], Haiwang Zhong[17], Taochun Sun[1], Hans Joachim Schellnhuber[18]

1 Department of Earth System Science, Tsinghua University, Beijing 100084, China.
2 Laboratoire des Sciences du Climat et de l'Environnement LSCE, Orme de Merisiers 91191 Gif-sur-Yvette, France
3 Department of Earth System Science, University of California, Irvine, 3232 Croul Hall, Irvine, CA 92697-3100, USA
4 CICERO Center for International Climate Research, Oslo, Norway
5 Institute Pierre-Simon Laplace, Sorbonne Université/CNRS, Paris, France
6 Kayrros, 33 Rue La Fayette, FR 75009, Paris, France
7 Global Carbon Project, CSIRO Oceans and Atmosphere, Australia Capital Territory, Australia
[8] College of Engineering, Mathematics and Physical Sciences, University of Exeter, Exeter, UK
[9]Department of Earth & Environment Sciences, Columbia University
[10] Earth System Science Department, Woods Institute for the Environment, and Precourt Institute for Energy. Stanford University, Stanford, CA, USA
[11]Energy and Resources Group and Goldman School of Public Policy, University of California, Berkeley, CA, USA
[12]School of Environmental Science and the Tyndall Centre for Climate Change Research, University of East Anglia, Norwich, UK
[13]European Commission, Joint Research Centre (JRC), Ispra, Italy
[14] Center for Global Environmental Research, National Institute for Environmental Studies, Tsukuba, Japan
[15] Key Laboratory of Land Surface Pattern and Simulation, Institute of Geographical Sciences and Natural Resources Research, Chinese Academy of Sciences, Beijing, China
[16] Tsinghua Shenzhen International Graduate School, Tsinghua University, Shenzheng, China
[17] Department of Electrical Engineering, Tsinghua University, Beijing 100084, China
[18]Potsdam Institute for Climate Impact Research, 14412 Potsdam, Germany

† Authors contributed equally

*Corresponding Author: zhuliu@tsinghua.edu.cn

**Abstract.** The diurnal cycle CO$_2$ emissions from fossil fuel combustion and cement production reflect seasonality, weather conditions, working days, and more recently the impact of the COVID-19 pandemic. Here, for the first time we provide a daily CO2 emission dataset for the whole year of 2020 calculated from inventory and near-real-time activity data (called Carbon Monitor project: https://carbonmonitor.org) for power generation (for 29 countries), industry (for 73 countries), road transportation (for 406 cities), aviation and maritime transportation and residential fuel use sectors (which we estimate for 206 countries). It was previously suggested from preliminary estimates that did not cover the entire year of 2020 that the pandemics may have caused more than 8% annual decline of global CO2 emissions. Here we show from detailed estimates of the full year data that the global reduction was only 5.4% (-1,901 MtCO2, ±7.2% for 2-sigma range). This decrease is 5 times larger than the annual emission drop at the peak of the 2008 Global Financial Crisis. However, global CO2 emissions gradually recovered towards 2019 levels from late April with global partial re-opening. More importantly, global CO2 emissions even increased slightly by +0.9% in December 2020 compared with 2019, indicating the trends of rebound of global emissions. Later waves of COVID-19 infections in late 2020 and corresponding lockdowns have caused further CO2 emissions reductions particularly in western countries, but to a much smaller extent than the declines in the first wave. Despite partial mobility restrictions in European countries and in some US States that have been affected by the second and third waves of the pandemics, emission reductions observed in Autumn 2020 were much smaller than those of Spring 2020. That even substantial world-wide lockdowns of activity led to a one-time decline in global CO2 emissions of only 5.4% in one year highlights the significant challenges for climate change mitigation that we face in the post-COVID era. These declines are significant, but will be quickly overtaken with new emissions unless the COVID-19 crisis is utilized as a break-point with our fossil-fuel trajectory, notably through policies that make the COVID-19 recovery an opportunity to green national energy and development plans.

**1 Background**

Emissions of CO2 from fossil fuel combustion and cement production ("fossil CO2") are the main cause of climate change (Friedlingstein et al., 2020). With decades of development of inventories, fossil CO2 emissions have been estimated by activity data (e.g., the amount of fossil energy consumption) and emission factors (e.g., the amount of CO2 emission per unit of fossil energy consumed). The uncertainty (range from ±6% to ±10%) of global fossil CO2 emissions is in general much lower than for other species such as air pollutants (Olivier and Peters, 2019). Current satellite observations do not allow us to verify national inventories and narrow down this uncertainty as they are too uncertain and influenced by natural sources or sink of CO2. Yet, satellite observations have providedconsistency checks with inventories for few large point sources (Nassar et al., 2017) and for point sources and cities (Zheng et al., 2020a). In response to the lockdown restrictions to prevent the spread of coronavirus in all countries (International Monetary Fund (IMF), 2021), the COVID-19 pandemic has had huge impacts on human activities and on the earth system (Diesendorf, 2020; Diffenbaugh et al., 2020; Klenert et al., 2020), leading to temporary declines in air pollutants (NOx, PM$_{2.5}$, SO$_2$ and etc.) (Bauwens et al., 2020; Huang et al., 2020; Le et al., 2020; Shi and Brasseur, 2020; Zheng et al., 2020b) and CO2 emissions (Forster et al., 2020; IEA, 2020b; Le Quéré et al., 2020; Liu et al., 2020a; Liu et al., 2020b).

Given the apparent lack of low-latency, direct energy-use related activity data for estimating global CO2 emissions, the effect of the pandemic on CO2 emissions can be estimated by combining those activity data and the change of activities through time. For example, Le Quéré et al. (2020) estimated the CO2 decrease during three levels of forced confinement based on the Oxford index of government response to the pandemics linearly scaled by activity data, and Forster et al. (2020) estimated GHG emission declines in 2020 using Apple mobility change data. With various proxy data available for different sectors, at daily or even hourly or sub-hourly frequency, it becomes possible to take the next step and build a near-real-time CO2 emission dataset at high temporal resolution.

In a recent study, Liu et al. (2020a; 2020b) described the Carbon Monitor fossil fuel CO2 emission dataset, which provide daily data for 6 sectors and 12 large emitting countries plus the rest of the world as an aggregate, up until December 31$^{st}$, 2020. The global fossil CO$_2$ emissions was separated into sectors of power generation (~40% globally), industrial production (~30%), transportation (~20%, ground, air and shipping) and residential consumption (~10%). Based on this methodology, the Carbon Monitor CO$_2$ emission dataset with daily resolution is updated in this study for the entire year 2020. This product is evaluated against preliminary national energy use data for all or part of the year 2020, providing a full picture of all the drivers of CO$_2$ emissions including the pandemic (seasonality, working days and holidays, weather and the economy). Acknowledging higher uncertainties than inventories – such a dataset provides more up-to-date information than official inventories (UNFCCC, 2020a, b, c) and international CO2 emissions datasets (BP, 2020; Crippa, 2020; Gilfillan et al., 2020; IEA, 2020a; Friedlingstein et al., 2020) which have a time lag of between six and 16 months after the last month of reported emission.

In this paper we provide a detailed description of the datasets and methodology used to estimate daily CO2 emissions for 11 countries/regions and the global total for the year 2020, covering the sectors of power generation, industrial production, transportation (ground, air and shipping) and residential consumption. Daily CO2 emissions in 2019 are used as reference data to compare the difference with emissions in 2020, which are caused by factors such as weather, temperature, industrial development and – most importantly – the impact from COVID-19 pandemic starting in early 2020. Additional materials associated with the release of each new version will be posted at the Carbon Monitor website (https://carbonmonitor.org).

## 2 Result

**Global daily CO2 emissions for 2020**

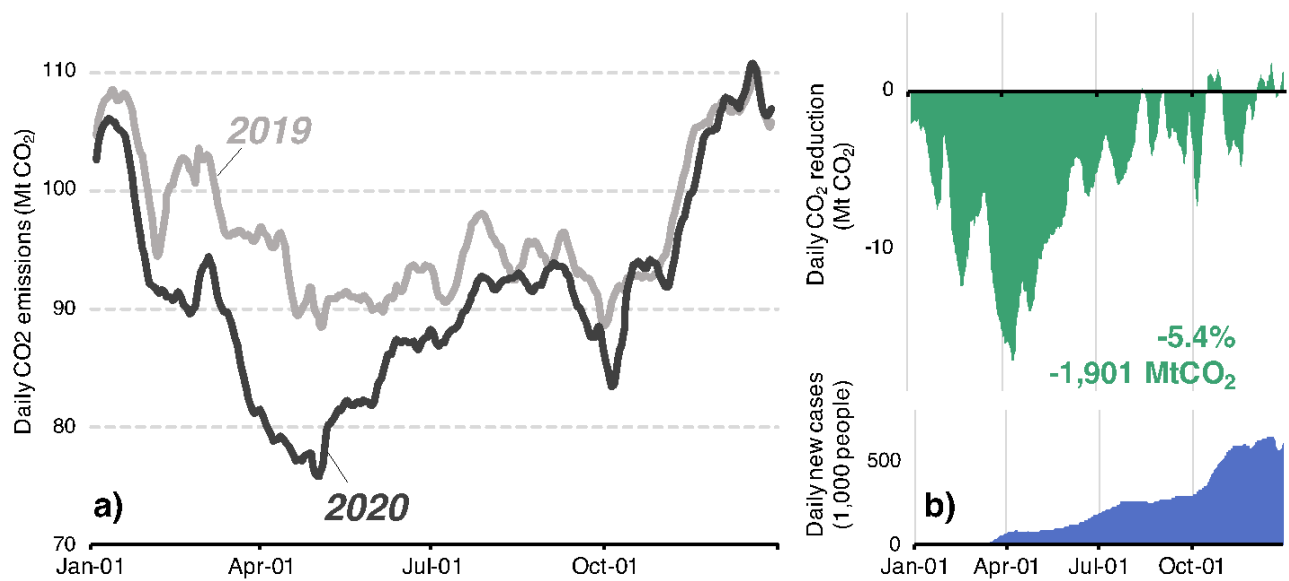

**Figure 1 | a) Global daily CO2 emissions in 2019 and 2020; b) global daily CO2 reduction in 2020 compared to 2019 and the global daily new cases of COVID-19 in 2020.** The dark black line shows the 7-day running mean of daily CO2 emissions in 2020 and results in 2019 are shown by a light black line.

The total global daily CO2 emissions for 2020 is 33.5 GtCO2, Fig 1, with power accounting for 39%, ground transportation 17%, industry 30%, residential 11%, domestic aviation 1% and international bunkers (international aviation and shipping) 2%. For the top five emitters in 2020, China's emissions in 2020 were 10.5 GtCO2 (power: 44%, ground transport: 8%, industry: 40%, residential: 8%, domestic aviation: 1%), followed by the USA at 4.6 GtCO2 (power: 31%, ground transport: 33%, industry: 21%, residential: 12%, domestic aviation: 3%), EU27 & UK at 2.9 GtCO2 (power: 29%, ground transport: 29%, industry: 20%, residential: 22%, domestic aviation: 0.3%), India at 2.3 GtCO2 (power: 51%, ground transport: 11%, industry: 29%, residential: 9%, domestic aviation: 0.2%), and Russia at 1.5 GtCO2 (power: 54%, ground transport: 15%, industry: 19%, residential: 1%, domestic aviation: 11.%).

Compared to 2019, the global CO2 emissions in 2020 decreased by an estimated -1,901 MtCO2, which represents a relative change of -5.4% (-1,901 MtCO2, ±7.2% for 2-sigma range) compared to 2019 (Fig 2a). In comparison, CO2 emissions were continuously growing with an average growth rate of 2.8 % $yr^{-1}$ in the 1970s (1970–1979), 1.3 % $yr^{-1}$ in the 1980s (1980–1989), 1.1 % $yr^{-1}$ in the 1990s (1990–1999), 2.7 % $yr^{-1}$ in the 1990s (2000–2009) and 1.6 % $yr^{-1}$ for the last decade (2010–

2019)[24]. The decrease of CO2 emission in 2020 (-1,901 MtCO2) is the largest ever absolute annual decline in emissions, larger than the emission decreases of 2009 financial crisis (-380 MtCO2) [24] and even than the decrease reconstructed in the end of World War II (-814 MtCO2)[23]. Specifically, we estimate emissions fell by -9.4% in USA, -9.8% in Brazil, -8.1% in India, -7.5% in EU&UK (UK: -9.5%, France: -9.0%, Germany: -7.9%, Italy: -7.4%, Spain: -13.1%), -5.0% in Japan, -2.9% in Russia. Conversely, China's CO2 emissions in 2020 increased slightly by +0.5%, because China reopened its economy earlier than most other countries. The significant decrease of CO2 emission is linked to the impact of complex responses to the COVID-19 pandemic, including stay home orders, closure of factories, collapse of air traffic and perturbations of supply chains. The largest weekly decline was found in Week 15, 2020 (April 6th – April 12th) by -18% in 2020 compared to the same week in 2019. The emission decline in the ground transportation sector contributed more than one third (37%) of the total global emission decline in 2020 compared to 2019.

Mean daily emissions were 91.4 $MtCO_2$ per day in 2020, which is 10% lower than the daily average emissions in 2019 (96.9 $MtCO_2$ per day). Importantly, although significant declines of CO2 emissions have been observed in 2020 that can mainly be attributed to the impact of the pandemic, other effects may have played a role such as prevailing warmer winter temperatures over most northern industrialized regions during the first months of 2020. In addition, global CO2 emissions gradually recovered from late April with global partial re-opening. Second and third waves of pandemics in Autumn and early winter 2020 and the corresponding new lockdowns reduced CO2 emissions further in western countries, but to a much less extent than the declines in the first wave. Global emissions were strongly reduced by the first wave of COVID-19, dropping most pronouncedly in April by -15.8% compared to the same month in 2019. However, although hit by further waves of infections in many countries, global CO2 emissions only dropped by -3.1% in November and even increased slightly by +0.9% in December. Decreases in mobility-related emissions seem to be more persistent than decreases from other sectors: emissions from ground transportation were -10.9% lower in 2020 than in 2019, with the largest monthly decreases occurring in April and May (-33.7% and -26.2% respectively) while monthly declines were much smaller in November and December (-9.7% and -6.8%). Emissions from the power sector and the industry sector decreased most significantly in April by -10.0% and -9.9% respectively but recovered to their 2019 levels from August, by the average growth rates of 1.0% and 2.5% respectively during August to December. However, emissions still decreased cumulatively by -2.5% and -1.4% for the whole year of 2020 compared to 2019 in the power sector and the industry sector.

**Daily CO2 emissions from major countries/regions in 2020**

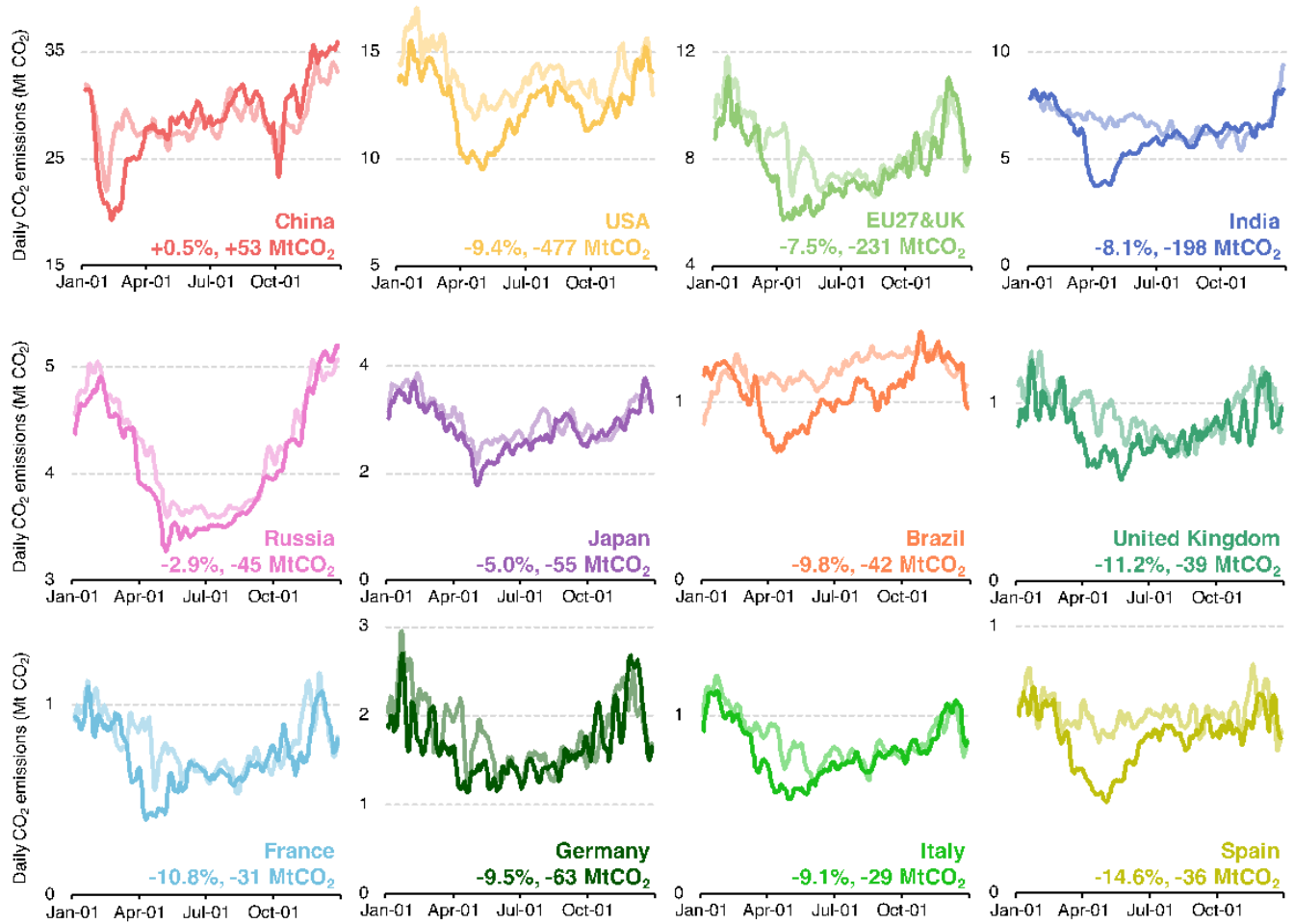

**Figure 2 | Daily CO2 emissions in 2019 and 2020 for countries.** The thick line shows the 7-day running mean of daily CO2 emissions in 2020 and results in 2019 are shown by a thin line.

**China**

China's CO2 emissions increased slightly by +0.5% (+53 MtCO2) in 2020 (Fig 2). The largest emission decline in China occurred in the first quarter of 2020 with a cumulative decline by -10.6% (-260 MtCO2), corresponding to the early outbreak of COVID-19 and strict lockdown measures, with the most pronounced weekly decline of -29.0% in Week 8, 2020 (February 17th – February 23rd). With the lift of lockdown from early April, China's CO2 emissions increased as social and economic activities gradually resumed. Monthly emissions showed a net increase since April, with an average increase of +3.9% (ranging from 0.4% to 7.9%) per month. The net increase in China was mainly driven by the rebound of the power sector (average increase averagely by +4.2% per month) and the industry sector (average increase +6.2% per month), as these two sectors account for more than 80% of total CO2 emissions in China. Since September and November 2020, respectively, China's cumulative emissions of the industry sector and the power sector have even surpassed the 2019 level.

**India**

India's emission decreased by -8.1% (-198 MtCO2) in 2020. Prior to COVID-19, India showed increased carbon emissions of 2.8% in January and 6.4% in February. The strict lock-down measures led to an immediate decline of carbon emissions, declining by -17.6% and -43.6% in March and April respectively. The most pronounced weekly decline was found in Week 15, 2020 (April 6$^{th}$ – April 12$^{th}$) with -45.5%. Emissions have steadily recovered since the middle of May 2020, regaining the 2019 level in September with a monthly increase by +4.1% and even greater since then, although this is largely due to very low emissions in the latter half of 2019 for other reasons (Andrew, 2020). However, the cumulative emission in India in 2020 was still lower than 2019, declining by -11.3% in the first three quarters of 2020 and by -8.1% in the whole year of 2020 compared to the same period of 2019.

**United States of America**

The total emission from the USA in 2020 decreased by an estimated -9.4% (-477.4 MtCO$_2$). Before March 2020, emissions were lower than in the same period in 2019, by -9.5% in January and -3.8% in February. The large drop of CO2 emissions from March coincides with the sudden increase of newly confirmed cases in the country. From March to June 2020, US CO2 emissions declined by -12.7%, -19.7%, -19.2% and -10.2% respectively. The most pronounced weekly decline was found in Week 18, 2020 (April 27$^{th}$ – May 3$^{rd}$) by -26.3%. With no new nation-wide stay-at-home order issued since the first wave, the emission declined showed little correlation with the second and third peak of COVID-19 active cases (peaked in July and December, respectively). In November, several state governments ordered regional stay-at-home recommendations, which led to a notable CO2 emission decline (-11.8% in November 2020). This indicates that CO2 emissions are more related to regional or national stay-at-home policies rather than with active case numbers. Christmas and New Year's holiday led to a pause in the decline in US CO2 emissions. CO2 emissions increased +6.3% from 24$^{th}$ December to 31$^{st}$ December, while the monthly emissions in December were back to the same level as 2019.

**EU&UK**

Total emissions from EU&UK in 2020 decreased by -7.5% (-231 MtCO2). EU&UK already showed CO2 emission declines prior to the pandemic. In January and February 2020: CO2 emissions from EU&UK declined by -8.7% and -4.7% compared to the same period in 2019, mainly attributable to the warmer temperature and lower power demand[25]. Under the impact of the pandemics and stay-at-home orders after mid March 2020, CO2 emissions in EU&UK declined by -8.1%, -26.3% and -21.6% for March, April and May, respectively. The most pronounced weekly decline was found in Week 15, 2020 (April 6$^{th}$ – April 12$^{th}$) with -37.4%. From June to September, with relaxed lock-down measures, emissions showed a smaller decline (average decline -3.3% per month) than during the first wave of infections (the average monthly emission decline from March to May was -18.7% per month. September was the first month in 2020 when emissions showed a slight increase (+0.5% in September 2020). Along with the impact of the second wave and subsequent stay-at-home orders, October and November again showed emissions decreases, of -4.5% and -6.5%, respectively. However, emissions in December 2020 showed a net increase of +3.5%, with larger monthly power emissions in December 2020 with a year-over-year growth rate of +11.8%, more than offsetting COVID-19 impacts.

European countries affected by second and third waves of the pandemics had responses of different severity (Fig 3) – UK, France, Germany and Italy – but they show similar emission drops by the end of the year 2020:

In the UK, there was a fall of -9.5% in 2020 compared to 2019 emissions and the peak of the monthly decrease in emissions (-28.8%) occurred in April during the first round of lockdown. This first COVID-19 lockdown hit every sector, with the ground-transport sector appearing to be the most affected, with emissions below their average 2019 levels by -39.7% for April and -30.7% for May. Emission rebounded in subsequent months and a net emission increase was found in August (+2.0%) and September (+3.1%) due to strong emission increase in power in these two months (21.7%; 18.0%). During the second round of lockdown, there is a clear emission drop in November (-13.9%) but much less significant emission decline compared to the first round. However, emissions in December increased +0.6% due to the strong monthly increase from the power sector (+6.9%) and residential sector (+4.2%).

In France, the national CO2 emissions dropped -9.0% in 2020 compared to 2019. Small reductions (-4.3%) were already seen in the first two months of 2020 and the first round of (very strict) national lockdown sharply further decreased emissions in April (-41.0%) and May (-26.8%). Emissions rose up to a net increase in September (1.4%) and October (0.9%) following a relaxing of lockdown restrictions. During the second wave, restrictions from Oct. 30 to Dec. 15, 2020 were not as strict as during the first one and we observed a noticeable but relatively small emission decline in November (-15.6%) and a slight increase in December (+0.3%). For 2020 as a whole, emissions in France decreased by -9.0% (-26 MtCO2).

In contrast, the expected reductions in emissions for 2020 were much more modest in Germany (-7.9%, -52 MtCO2) and Italy (-7.4%, -23 MtCO2). For Germany, the small emission decline is due to less strict lockdowns. For example, emissions declined in Germany by -27.2% and -21.5% in April and May respectively, the smallest emission decline during the first lockdown among the major European countries. At the same time, even though a second lockdown in Germany began in early November, there was only a limited emission reduction of -5.0% and an overshoot of +8.0% in November and December. Unlike other countries, in Germany, emissions from ground transport – a major source – barely changed in response to lockdowns while emissions from other sectors declined comparably, therefore the total emission drops due to the pandemic and for the whole year were smaller than for other large emitters in Europe. As for Italy, the smaller relative emission decrease for the year 2020 may come from a late second lockdown. Emissions had been cut dramatically by -29.5% and -25.8% in April and May under the impact of the first round of (strict) lockdown.

For Spain, which did not have a second round of mandatory lockdowns, there was a -13.1% (-32 MtCO2) decrease in emissions in 2020 as a result of extensive lockdown measures introduced in March, along with the impact of the second wave of infections observed in October (-9.4%) and November (-9.0%).

**Russia**

CO2 emissions in Russia decreased by -2.9% (-45 MtCO2) in 2020 compared to 2019, with the largest weekly decline in Week 14, 2020 (March 30th – April 5th) at -9.7%. Prior to COVID-19, Russia showed emission decline in January by -3.6 %, corresponding to the warmer temperature compared to 2019, and a small increase in February by +0.9%. Under the impact of the pandemic's first wave, emissions in Russia decreased by -2.5% in March, -7.7% in April and May, and -5.3% in June respectively. The emissions in the following months were still lower than the same month in 2019, until December 2020, when Russia's emissions increased by +3.0%.

**Japan**

CO2 emissions in Japan decreased by -5.0% (-55 MtCO2) in 2020 compared to 2019, with the largest weekly decline in Week 19, 2020 (Mar 4th – Mar 10th) by -24.8%. Japan already showed reduced CO2 emissions prior to the pandemic, which slowly intensified from March. CO2 emissions declined by -5.6% and -2.2% in January and February, respectively. The gradually introduced stay-at-home orders led to further declines in emissions. The biggest monthly emission decline was in May, -17.5% lower than in May 2019, corresponding to the peak of the pandemic's first wave in Japan. However, the second wave (peaked in August) and third wave (intensified from November to the end of 2020) did not cause as much emission decline.

**Brazil**

CO2 emissions in Brazil decreased by -9.8% (-42 MtCO2) in 2020 compared to 2019, with the largest weekly decline in Week 15, 2020 (April 6th – April 12th) by -35.7%. From March, under the impact of COVID-19, Brazil showed large emission declines compared to 2019. The biggest monthly decline occurred in April 2020, decreasing by -31.0% compared to the same month in 2019. The emissions were decreasing in the following months of 2020; however, the declines have been smaller since September 2020 (dropping by -14.4% in September, -4.8% in October, -0.8% in November). In December 2020, the CO2 emission in Brazil even increased slightly by +2.3%.

**Daily CO2 emissions by sectors in 2020**

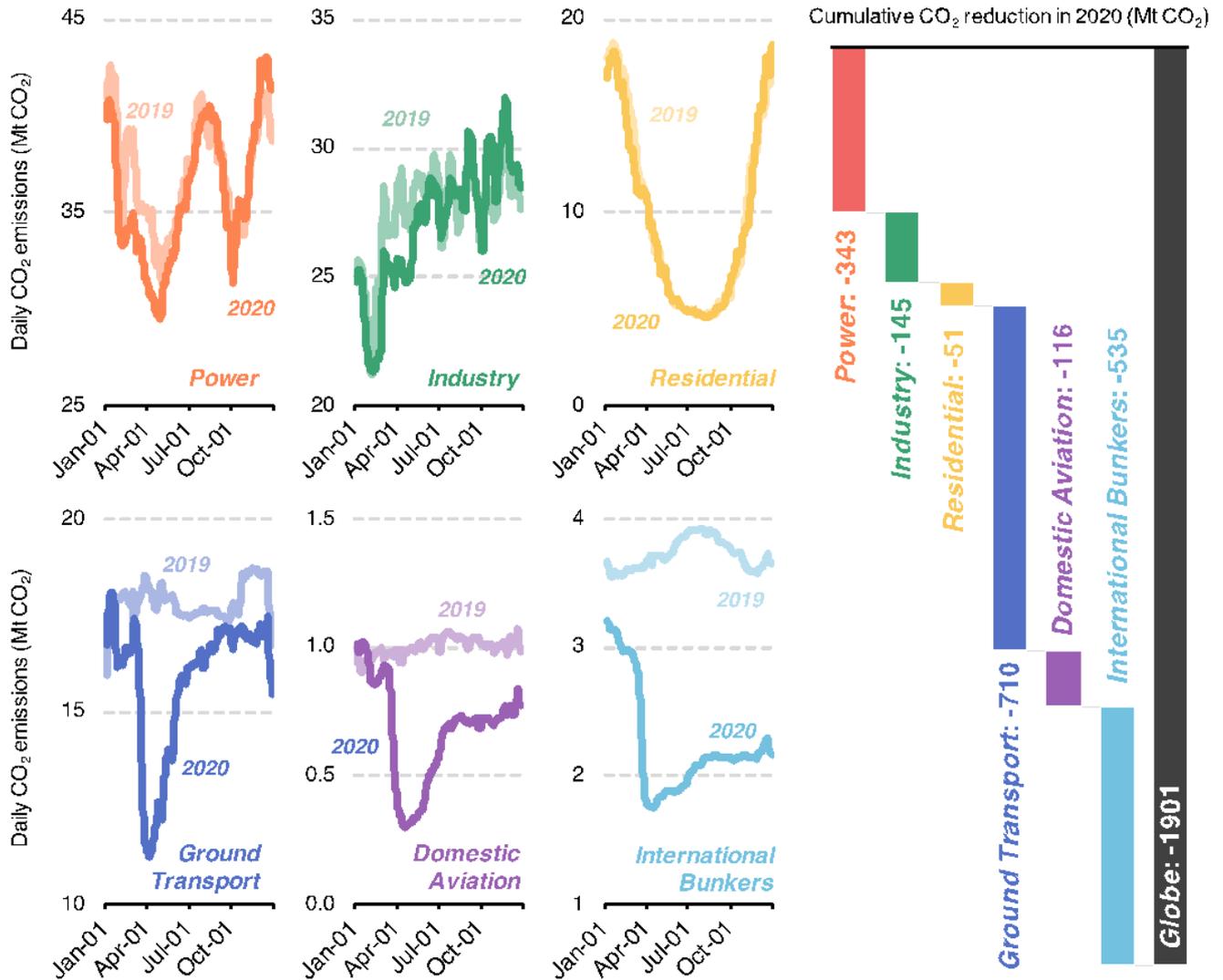

**Figure 3 | Global daily CO2 emissions by sectors in 2019 and 2020.**

The global decrease in CO2 emissions is mainly due to the mobility-related emissions, with the largest contributions to the global decrease in emissions in 2020 coming from ground transportation (-947 MtCO$_2$, down -14.6% and 47% of the total decrease) and the aviation sector (including domestic aviation and international aviation, -466 Mt CO$_2$, down 47.6% and 23% of the total decrease). Somewhat smaller decreases were observed from the power sector (-272 Mt CO$_2$, 13% of the total decrease) and the industry sector (-140 Mt CO$_2$, 7% of the total decrease), and relatively small decreases in international shipping (-181 Mt CO$_2$, 24.8% of the total decrease) and in residential sector emissions (-29 Mt CO2, 1% of the total decrease). Further details of these sectoral changes (Fig. 4) are discussed below.

The global decrease is mainly due to the mobility-related emissions, such as the ground transportation sector (-10.9%), the domestic aviation sector (-31.9%), and the international bunkers (-39.4%). The largest contributions to the global decrease in emissions in 2020 come from ground transportation (-710 MtCO$_2$, 37% of the total decrease) and the international bunkers (including international aviation and international shipping, -535 Mt CO$_2$, 28% of the total decrease), with somewhat smaller decreases from the power sector (-343 Mt CO$_2$, 18% of the total decrease), the industry sector (-145 Mt CO$_2$, 8% of the total decrease), the domestic aviation sector (-117 Mt CO$_2$, 6% of the total decrease), and relatively small decreases in residential sector (-51 Mt CO2, 3% of the total decrease). Further details of these sectoral changes (Fig. 3) are discussed below.

### 3.3.1 Power emissions

Fig 3 shows that for the year 2020 as a whole, global CO2 emissions from the power sector declined by -2.5% (-343 Mt CO2), with decreases in the USA (-10.3%, -166 MtCO2), India (-4.4%, -53 MtCO2), Brazil (-6.5%, -5 MtCO2) and EU&UK (-11.5%, -108 MtCO2) but a small increase in China (+1.2%, +55 MtCO2). Some of the drop in China's power-sector emissions are due to prevailing warmer winter temperatures, and the near-zero differences in late January and early February between 2020 and 2019 are because this was when the country's spring festival occurred in 2019 (Fig 3). Daily emissions rebounded rapidly after the Spring Festival in 2019, however, in 2020, the emissions increased slowly due to the pandemic impacts. At the end of March 2020, emissions were back to similar levels as in 2019; since November, the cumulative emissions in 2020 were even greater by 0.3% in the first 11 months and 1.2% in the whole year compared to the same period of 2019. In addition, the declines of power generation affected by the festivals such as Ching Ming Festival (April 5$^{th}$), the Labor Day holiday (May 1$^{st}$-5$^{th}$) and the National Day holiday (October 1$^{st}$-8$^{th}$) are also reflected by the data. Meanwhile, power emissions in Russia and Japan were almost stable in whole of 2020 (-3.1%, -26 MtCO2; -2.8%, -15 Mt CO2). It should be noted that power emissions in Spain were below their normal levels in 2019 all through the year, even before the pandemic hit. This low annual growth in emissions (-23.9%, -13 MtCO2) was a result of strong renewables generation.

**Industrial emissions**

Global industry emissions in 2020 fell by -1.4% (-145 MtCO2) in most countries, including USA (-5.5%, -56 Mt CO2), EU&UK (-8.2%, -52 Mt CO2) and India (-13.1%, -99 Mt CO2). In contrast, China's industrial emissions increased by +2.9% (+120 Mt CO2) in 2020. This surprising result shows the central position of industry in China's various economic activities. In China, despite the pandemic, industrial emissions only dropped by -6.4%, -16.8% and -8.0% in January, February, and March, respectively. Since April, industrial emissions have increased, and started to show an increase of +0.8% in the first nine months in whole year 2020, with a considerable rise of +8.1% in September, a result of recovery in various industrial activities. For most countries, after experiencing the most significant reduction in April, including the USA (-11.7%), EU&UK (-29.5%) and India (-66.6%), the reductions of their industry emissions have been shrinking in the remaining months of 2020 as well as the major holidays afterwards, suggesting that COVID-19 is the main driver of emission changes from industry sector.

**Residential**

Residential energy consumption is mostly driven by the temperature, which showed no significant change during the lockdown period (i.e., no significant change with regards to the year-to-year variability when the contribution of temperature variability was removed). Although increased time at home may increase emissions from commercial and residential consumption, however, abnormally warm northern-hemisphere winter and early spring in 2020[26,27] may likely have resulted in the decrease of heating demand and emissions. Combining the impacts of temperature and the COVID-19 pandemic, the emissions from global residential consumption only dropped slightly by -0.8% (-29 MtCO2) in the whole year 2020.

**Ground transportation emissions**

Emissions from ground transportation were calculated based on TomTom congestion level with daily transportation activity data for 416 global cities in more than 50 countries. The daily dynamics are sensitive enough to capture even a short-lived drop due to major holidays for most countries. Overall, emissions from ground transportation decreased dramatically by -10.9% in 2020 (-710 MtCO2) contributing the largest share of emission decline among all sectors (Fig 3). In China, as cities started locking down in the last week of January, average emissions from transport during that month decreased by -18.6%. In February, ground transport emissions dropped abruptly by -54%, compared to the same month in 2019. However, the reduction became smaller from March (-25%) to September (-0.3%). By September 2020, China's emissions were not different than those during the same period in 2019 (-0.3%). The decline became larger again from October, with emission reductions by -7%, -2% and -4% in October, November and December, respectively. In most other countries, emissions from ground transport dramatically decreased just after lockdown measures starting in the latter half of March, with the largest decline of the year occurring in April. Ground transport emissions in the USA and India dropped by -9% and -26% in March since the lockdown measures, with the largest declines in April with -30% and -66%, respectively, and continued to decrease in subsequent months. Emissions in European countries dropped starting in March (-16.7%), with the UK, France, Germany, Spain and Italy showing the largest reductions of -40%, -51%, -16%, -53% and -40% respectively in April in 2020 (Fig 14). For most countries, the biggest drop occurred in April, with less reduction in subsequent months. Except at the end of 2020, especially in November, the drop in emissions increased again due to the partial recurrence of the pandemic, including for India (-8.0%), USA (-6.9%), and EU & UK (-10.0%).

**Aviation and shipping emissions**

Emissions from global aviation decreased by -47.6% in 2020 (-466 MtCO2) (Fig 15), among which those from international aviation decreased by -354 MtCO2. The aviation sector is the sector with the largest decline rate of emissions for the year as a whole, with emissions from domestic aviation and international aviation dropping by -31.9% and -56.4%, respectively.

International aircraft emissions are included in our global estimates, but only domestic aviation emissions are attributed to countries, following standard emissions accounting practice. The total number of flights and global aviation emissions shows two consecutive decreases, one by the end of January in Asia, and another since the middle of March in the rest of the world, which lasted through the end of 2020. Emissions declined sharply from mid-March, coincident with travel bans and lockdown measures. CO2 emissions from international shipping decreased by -24.8% (-181 MtCO2) in 2020, although this estimate is less certain than those for other sectors.

**Diminished impacts on emission reduction during the second national lockdown period**

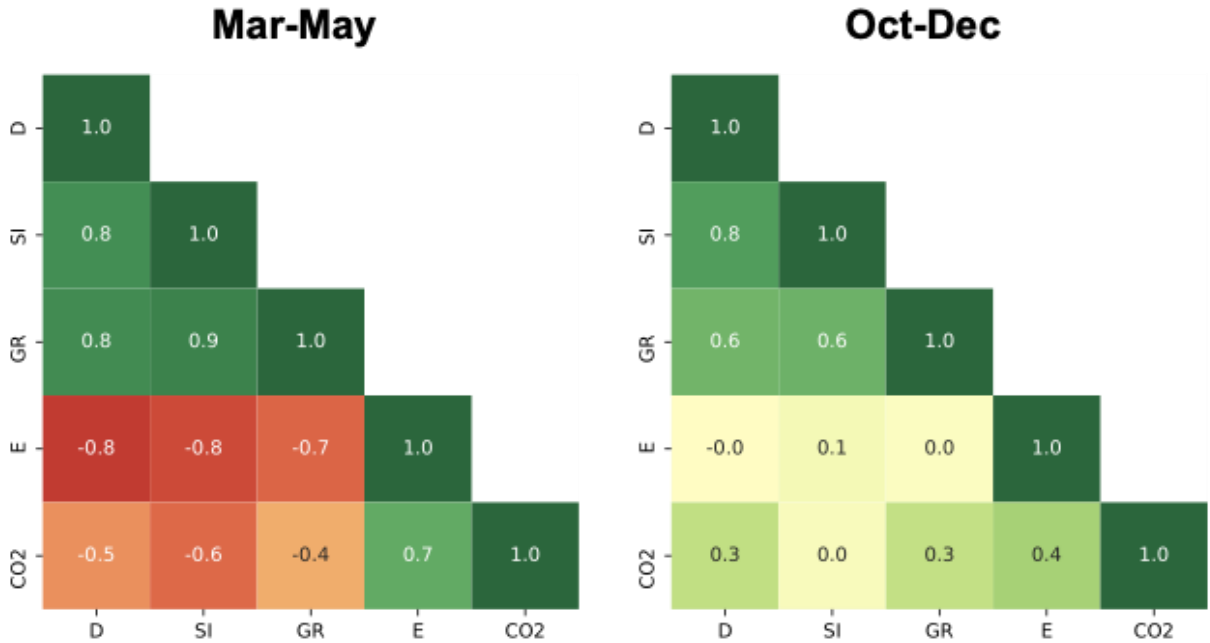

Fig 4 | **Correlation matrices of five indicators during March 1$^{st}$ ~ May 31$^{st}$, 2020 and October 1$^{st}$ ~ December 31$^{st}$, 2020.** $D$ denotes the daily deaths of COVID-19, $SI$ denotes the stringency index of government responses to COVID-19, $GR$ denotes the duration spent at places of residence, $E$ denotes the daily changes of power demand in 2020 compared to the same day in 2019, and $CO2$ denotes the daily CO2 changes in 2020 compared to the same day in 2019. These five indicators are the averages of U.S., India, UK, France, Germany, Italy, Spain, Russia, Brazil, and Japan (country-specific correlation matrixes are shown in SI Fig 2).

Compared to March to May, 2020, there were similar numbers of deaths of COVID-19, but less CO2 emissions reduction during October to December, 2020. Daily deaths in the U.S., India, UK, France, Germany, Italy, Spain, Russia, Brazil, and Japan, increased averagely by around 3,120 during March to Mar, 2020, while around 4,694 during October to December, 2020. However, their total CO2 emissions dropped by -6 $MtCO_2$ per day, compared to only -1 $MtCO_2$ per day during the last three months of 2020.

The dramatic drop of CO2 during March to May 2020, showed strong correlation to the numebr of daily deaths of COVID-19 as well as the level of government response, human mobility, and energy demand. To prevent the spread and reduce deaths, governments tightened their policies and even implemented closure measures regionally or nationwide. Accordingly, lives and livelihoods were confined to some extent as people spent more time at their residences, and the demand for energy was reduced. Thus, the duration people spent at home and the reduction of energy demand show very strong relationship with the stringency index of government response, for example, longer duration people spent at home and larger reduction of energy demand along with the harsher government measures. As a result, less human activities resulted in a significant drop of CO2 during this period.

However, during October to December 2020, the changes of human mobility and energy demand show weaker or even no relationship with the stringency of government response, although governments still maintained the high level of confinement measures to cope with the new wave of spread and deaths of COVID-19. It could be explained that as people resumed their lives and livelihoods to a limited extent with a deeper understanding of the COVID-19, and governments partially restored their economy, although governments were still maintaining high-pressure measures to respond to the pandemic. As the result, the relationship between the duration people spent at residences and the stringency index of government responses was weaker (r=0.6) compared to March-May 2020 (r=0.8), and the relationship between energy demand and the stringency index of government responses was even negligible (r=0.1) while it was strongly correlated during March-May 2020 (r=-0.8). However, the relationships between daily CO2 changes and duration at residences, between daily CO2 changes and energy demand were weaker during this period as well. Consequently, the daily CO2 changes show no relationship with the level of government responses.

In addition, the daily difference of temperature in 2020 compared to 2019 only resulted in a drop of -0.1% during March-May and an increase by +0.5% during October-December respectively.

Table 1 | **Daily average of the five indicators during the first and second waves of COVID-19.** $CO2$ denotes the daily CO2 changes in 2020 compared to the same day in 2019, $DC$ denotes the daily deaths of COVID-19[28], $SI$ denotes the stringency index of government responses to COVID-19[29], $GR$ denotes the duration spent at places of residence[30], and $E$ denotes the daily changes of power demand in 2020 compared to the same day in 2019.

|  | CO2 (Mt CO2/day) | DC (person/day) | SI (%) | GR (%/day) | E |
| --- | --- | --- | --- | --- | --- |

|  |  |  |  |  | (GWh/day) |
|---|---|---|---|---|---|
| March-May, 2020 | -6.06 | 3,120 | 67.89 | 15.65 | -1877.64 |
| October-December, 2020 | -0.96 | 4,694 | 62.60 | 8.65 | -11.26 |

**3 Comparison with Lequéré et al. estimates**

Lequéré et al. (LQ) published estimates of emissions reduction due to the covid crisis, with a methodology based on confinement indices and activity data. They established a correspondence between activity reductions and confinement severity changes, and then used daily confinement severity time series to produce daily emissions. A linear relation was assumed between relative changes of activity and emissions in each sector. By design, this approach provides an estimate of changes of emissions attributed to the pandemic and ignores other factors such as a higher renewable production, lower gas market prices, and lower heating demand from warmer cold season temperatures during 2020 relative to 2019. Changes in the methodology of LQ were brought between the first publication and the emissions projection to 2020 as described by Friedlingstein et al. 2020. The projection assumed that countries where confinement measures were at level 1 (targeted measures) on 13 November remained at that level until the end of 2020, a scenario that was too optimistic as new lockdowns happened in European countries and the US after November. Fig. 4 displays the comparison between LQ updated to cover the full year of 2020 as in Friedlingstein et al. 2020, and the Carbon Monitor (CM) estimates from this study. Comparison with other preliminary emission estimates for 2020 based on energy data are given in SI.

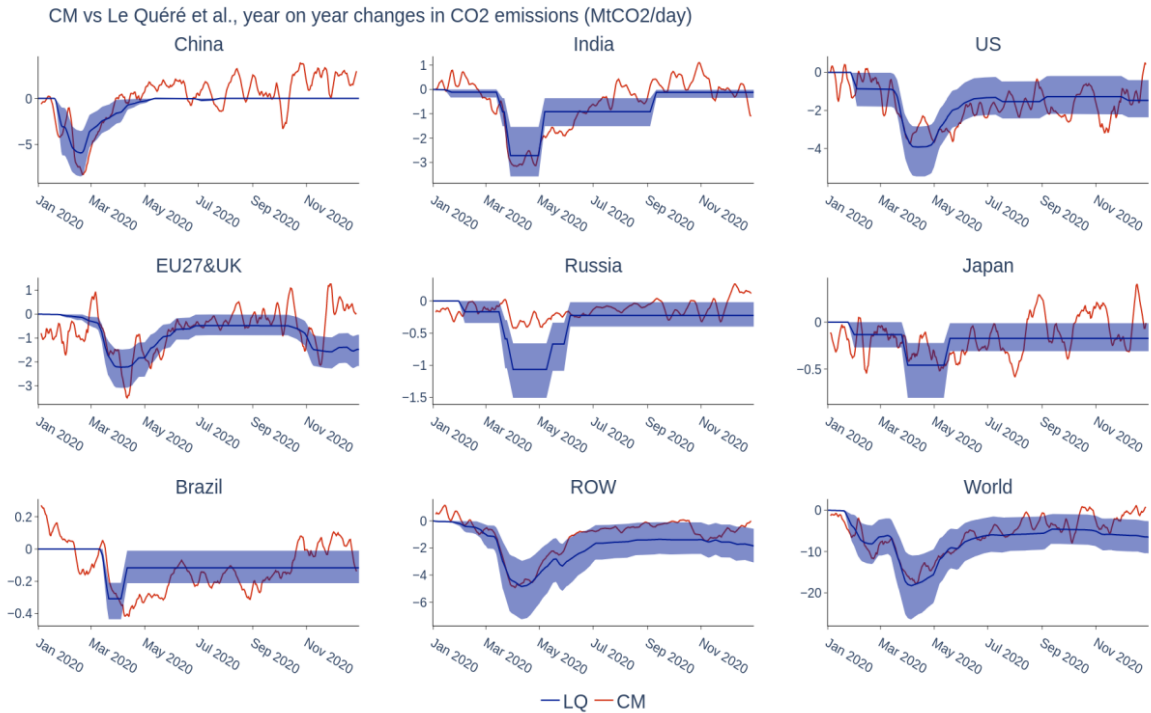

Fig 5: Comparison of year on year emission changes (2020 minus 2019) between Carbon Monitor (this study) and LQ by country. The Carbon Monitor daily data were applied a 7-days smoothing filter

During the first batch of lockdowns, confinement severity calibrated emissions changes from LQ produce CO2 emissions relative reductions consistent with CM, although more smoothed in time because of the discrete changes of confinement indexes. This is not surprising since many source activity data are common to both studies, except in Russia and Brazil where CM used specific regional data, hence the differences in Fig 5 for these two countries. During the second half of 2020 and the second wave of covid which saw lockdowns and restrictions reappear in some countries, CM and LQ show significant differences. The CM data indicate that $CO_2$ emissions went down by only 2% in the last quarter of 2020, compared to a 5.6% reduction in LQ (with the same CM2019 baseline). The smaller CM reduction based on actual activity data and accounting for the second batch of lockdowns suggests a degree of adaptation of human activities to the second wave which maintained more emissions. This indicates that the relationship between confinement severity and emissions was not stationary between the first and second waves. Moreover, activity-based estimation of emission reductions can provide unexpected insights. In Brazil, the short lockdown period fails to show that the mobility reduction was extended well into May. In Russia, it seems that the lockdown period did not influence emissions at all. All these regional discrepancies will be investigated in detail in further work.

**4 Conclusion**

The results show a decline of -5.4% (-1,901 MtCO2, ±7.2% for 2-sigma uncertainty) in global CO2 emissions in 2020 compared to 2019. The countries with the largest contributions to this global decline were the USA (-9.4%, -477 MtCO2), EU & UK (-7.5%, -231 MtCO2) and India (-8.3%, -198 MtCO2). Although China underwent the earliest lockdowns in response to the pandemic, China's emissions increased by +0.5% (+53 MtCO2) in 2020 owing to the relatively short duration of lockdowns. The total emission reduction in 2020 of -1,901 MtCO2, is about 5 times larger than the annual emissions decline at the peak of the Global Financial Crisis of 2008 (-380 MtCO2).

The significant emission decline is attributed to emissions decline from ground transport (-10.9%, -710 MtCO2), international aviation and shipping (-39.4%, -535 MtCO2), power generation (-2.5%, -343 MtCO2), industry (-1.4%, -145 MtCO2), domestic aviation (-31.9%, -117 MtCO2), and residential consumption (-1.4%, -51 MtCO2). The largest contribution to COVID-related decreases in emissions were from the ground transport sector (37%), followed by international aviation (28%), power generation (18%), international maritime transport (8%), industry (6%) and residential consumption (3%).

Our updated annual CO2 estimates for the full year of 2020 show a -5.4%±7.2% decline globally, which compares well with the multi-model median of 6.7% in the Global Carbon Budget 2020 using only partial data for 2020 [31]. Using partial 2020 data [31] different models showed decreases of -6.5% for the Carbon Monitor, -6.9% by Le Quéré, et al. [14], -13.0% by Forster, et al. [17], and 5.6% using the Global Carbon Project's standard method. In April 2020, the International Energy Agency (IEA) estimated that in the first quarter of 2020, global CO2 emissions had decreased by -5% compared to the first quarter of 2019, which were expected to decrease by -8% for the whole year of 2020 [13]. Recently, IEA revised its estimation for global CO2 emissions in 2020, with a drop of -6.8% for the whole year of 2020 [32]. In addition, our estimates for the annual CO2 decline in the USA is -9.4%, similar to the most recent estimate from the US Energy Information Administration (EIA) released in February 2021 with a whole year decrease of 2020 by -11.3% [33]. We also compared our estimates of daily CO2 reduction with study Le Quéré, et al. [14] (**SI Figure 1**).

Our correlation analysis shows the differing relationships between CO2 reduction and other indicators (daily new cases of COVID-19, stringency of confinement policy, duration people spent at home, and power demand) during the different periods of the first wave (March-Mar, 2020) and the second wave (October-December, 2020). During the first wave of the pandemic, governments generally adopted stricter confinement measures to prevent the spread of the coronavirus. Following the experience of the first lockdowns, with hard-hit economies subsequent lockdowns were less strict, and even though later waves caused wider spread of infection, governments and citizens resumed part or more of their activities. That reveals the urgent need to restore the economy which is also reflected by the difference in the degree of CO2 reduction in the different periods. In addition, in the post-COVID era, there are still huge conflicts between economic recovery and the control and reduction of greenhouse gas emissions.

Moreover, although the pandemic outbreak across the world is still underway, global CO2 emissions rebounded in the second half of 2020. There has been some discussion about possible "green" or "brown" recoveries from the emission decline in 2020 (Hepburn et al., 2020; Lahcen et al., 2020; Shan et al., 2020), considering the effects of fiscal stimulus packages in the post-COVID period and limiting increases in global average temperatures to 1.5 °C. However, current data availability are still not enough to fully capture the dynamics of CO2 emissions under the COVID-19 pandemic. Further monitoring, observation, data collecting and improved methods are urgently needed. The ability to monitor trends in daily emissions in near real time that we demonstrate here could contribute to timely policy actions with implications for climate change mitigation and earth system management.

**4 Methods**

We calculate daily CO2 emissions since January 2019, drawing on hourly datasets of electricity power production and CO2 emissions in 29 countries (including the substantial variations in carbon intensity associated with electricity production), three different indexes of daily vehicle traffic / mobility in 416 cities worldwide, monthly production data for cement, steel and other energy-intensive industrial products in 73 countries, daily aircraft transportation activity data, as well as proxies for the residential and the commercial building emissions.

**4.1 Daily emission estimates**

CO2 emissions can be estimated by multiplying the activity data (such as energy consumption) with their respective emission factors (CO2 emissions per unit of activity data)(IPCC, 2006; Liu et al., 2015):

$$Emis = \sum\sum\sum AD_{i,j} \times EF_{i,j} \qquad (1)$$

Here, $i$ and $j$ reflect the regions and sectors respectively. In our calculation, $i$ covers countries. $j$ covers six sectors that are power generation, industry, ground transportation, aviation, international shipping and household consumption. Due to data availability, we assume the emission factors remain unchanged during 2019 and 2020, thus, the daily emissions are directly proportional to the daily activity data.

$$\frac{Emis\prime}{Emis} = \frac{AD\prime}{AD} \qquad (2)$$

Specifically, the daily emissions in sector s were generally calculated by following steps: 1) we disaggregate the annual emissions in 2019 into daily level by following Eq. (3); 2) then we calculate the daily emissions in 2020 by the daily activity changes as Eq. (4), based on the aforementioned assumption of the linear relationship between activity data and emissions.

$$Emis_{d,2019} = \frac{AD_d}{AD_{2019}} \times Emis_{d,2019} \qquad (3)$$

$$Emis_{d,2020} = \frac{AD_{d,2020}}{AD_{d,2019}} \times Emis_{d,2019} \qquad (4)$$

Detailed methodologies have been discussed in our previous studies (Liu et al., 2020a; Liu et al., 2020b). However, in this study we update the baseline emissions in 2019 of each country and each sector based on the latest emission data release from EDGAR (Crippa, 2020). In addition, due to new data availability, we have updated our data sources and methodologies in some sectors. Full data sources have been listed in SI Table 1.

In the power sector, we followed Eq. (3) and Eq. (4), by using the daily national thermal production (or the daily total power production in Russia) to estimate the daily $CO_2$ emissions from the power sector. Since September 2020, we use the daily coal consumption of the Zhedian Company to disaggregate the monthly thermal generation data from China's National Bureau of Statistics (NBS), to estimate daily thermal generation in China. In addition, the OCCTO from Japan publishes hourly generation data with around 5-weeks' lag. Thus, we use the sum of the real-time daily electricity demand data from ten power companies in Japan to linearly predict the daily thermal production in December 2020.

In the industry sector, we primarily use the industrial production data or the industrial production index to calculate monthly emissions. But in some countries, due to the delay of data release by one month (China, USA Russia, and Japan) to two months (such as Brazil, India and European countries), we use the monthly prediction data of industrial production from the Trading Economics website (https://tradingeconomics.com/) to predict the changes of monthly $CO_2$ emissions. Firstly, we calculate the monthly emissions from the industry sector in 2019 by following the disaggregation Eq. (3) and then estimate the monthly emissions in 2020 based on the year-on-year rates of industrial production by following Eq. (4). Then we disaggregate monthly emissions using daily thermal electricity generation due to the lack of daily industrial data.

In the ground transportation sector, the activity data we used in this study (the traffic congestion level) were not directly proportional to emissions. However, the traffic congestion level is correlated with car counts, which is positively associated with emissions from ground transportation. Thus, we further develop a sigmoid model to describe the daily relationship between the congestion level and the car counts. Detailed information could be found in our previous paper (Liu et al., 2020a; Liu et al., 2020b).

In the aviation sector, we estimate both the domestic and international aviation emissions based on real-time flight distance (https://flightradar24.com). In the international shipping sector, our previous estimates were 25% according to a news report (Kinsey, 2020) of ship traffic volume decline in the first half year of 2020. Due to the lack of available real-time data, we retained the same estimation of decline by 25% in the whole year of 2020. Then, we calculate the daily average emissions as our estimates.

**U.S. daily emissions estimates**

Annual total state-level $CO_2$ emissions by sector in 2017 are obtained from U.S. Energy Information Administration (EIA; https://www.eia.gov/environment/emissions/state/), and then updated to 2018 based on EIA's latest comprehensive state-level annual estimates of energy consumption by sector and source (https://www.eia.gov/state/seds/seds-data-complete.php?sid=US). We disaggregate the annual emissions in 2018 into the monthly level for each sector using state-level monthly energy consumption data from EIA (SI Table 2). We also estimate monthly emissions by sector in 2019 and 2020 based on the change of monthly energy consumption data in 2019 and 2020 compared to the same period of 2018, by assuming that the emission factors remain unchanged. The monthly emissions are then allocated to each day using state-level daily indicators for each sector (SI Table 2). For the last two months in 2020, due to the lack of monthly energy data, we directly estimate daily emissions based on the change of daily indicators for each sector (SI Table 2), as well as scale factors that reflect potential change of carbon intensity of indicators in 2020 compared to the same period of 2019 (based on the change in previous month).

**4.2 Data sources**

The full data sources have been listed in Table 1. Compared to our previous releases, we have updated our data sources and methodologies in some sectors due to the data availability.

In the power sector, since September, 2020, we use the daily coal consumption of the Zhedian Company to disaggregate the monthly thermal production data from the National Bureau of Statistics, China (NBS), to estimate the daily thermal production in China. In addition, the OCCTO from Japan publishes hourly generation data with around 5-weeks lag. Thus, we use the sum of the real-time daily electricity demand data from ten power companies in Japan to linearly predict the daily thermal production in December, 2020.

In the industry sector, we primarily use the industrial production data or the industrial production index to calculate the monthly emissions. But in some countries, due to the delay of data release by one to two months, we use the monthly prediction data of industrial production from the Trading Economics website (https://tradingeconomics.com/) to predict the changes of monthly CO2 emissions.

In the international shipping sector, our previous estimates were 25% according to the news report of ship traffic volume decline in the first half year of 2020. Due to the lack of available data, we remained the same estimation of decline by 25% in the whole year off 2020.

**4.3 Uncertainty estimates**

Our uncertainty estimates for daily emissions are in addition to the uncertainties in annual emissions, which are as much as ±10%. We estimate the uncertainties for each sector separately, then aggregate them to estimate the combined uncertainty of the whole dataset. The overall uncertainty for the annual total of our dataset is ±7.2%. The two largest uncertainties come from the residential sector (40%), primarily coming from the underlying assumption of unchanged emissions from cooking as a result of incomplete residential consumption data. The high uncertainty in the industry sector (36%) is mainly attributable using daily thermal production to disaggregate the monthly industrial emissions to daily emissions.

We first calculate the uncertainties for each sector:
- Power sector: the uncertainty is mainly from inter-annual variability of coal emission factors and changes in mix of generation fuel in thermal production. The uncertainty of power emission from fossil fuel is within (±14%) with the consideration of both inter-annual variability of fossil fuel based on the UN statistics and the variability of the mix of generation fuel (the ratio of electricity produced by coal to thermal production).

- Industrial sector: The uncertainty of $CO_2$ from industry and cement production comes from monthly production data. $CO_2$ from industry and cement production in China accounts for more than 60% of world total industrial $CO_2$, and the uncertainty of emissions in China is 20%. Uncertainty from monthly statistics was derived from 10,000 Monte Carlo simulations to estimate a 68% confidence interval (1 sigma) for China. We calculated the 68% prediction interval of the linear regression models between emissions estimated from monthly statistics and official emissions obtained from annual statistics at the end of each year to deduce the one-sigma uncertainty involved when using monthly data to represent the change for the whole year. The squared correlation coefficients are within the range of 0.88 (e.g., coal production) and 0.98 (e.g., energy import and export data), which indicates that only using the monthly data can explain 88% to 98% of the whole year's variation[3]; the remaining variation is not covered but reflects the uncertainty caused by the frequent revisions of China's statistical data after they are first published.
- Ground Transportation: The emissions from the ground transportation sector are estimated by assuming that the relative magnitude in car counts (and thus emissions) follow a similar relationship with TomTom congestion index in Paris. The uncertainty in emissions were quantified by the prediction interval of the regression. Applying such a regression to all the 416 cities across the world might introduce additional uncertainties when other cities have a different relationship between  and TomTom congestion level, but this uncertainty is not quantified in this study due to the lack of similar car counts data for a wide range of cities across different countries.
- Aviation: The uncertainty in the aviation sector comes from the difference in daily emission data estimated based on the two methods. We calculate the average difference between the daily emission results estimated based on the flight route distance and the number of flights and then divide the average difference by the average daily emissions estimated by the two methods to obtain the uncertainty in $CO_2$ from the aviation sector.
- Shipping: We used the uncertainty analysis from IMO as our uncertainty estimate for shipping emissions. According to the Third IMO Greenhouse Gas study 2014[38], the uncertainty in shipping emissions was 13% based on bottom-up estimates.
- Residential: The 2-sigma uncertainty in daily emissions is estimated as 40%, which is calculated based on a comparison with daily residential emissions derived from real fuel consumption in several European countries, including France, Great Britain, Italy, Belgium, and Spain.

Then, we combine all the uncertainties by following the error propagation equation from the IPCC. Eq. 5 is used to derive the uncertainty of the sum and could be used to combine the uncertainties of all sectors:

$$U = \frac{\sqrt{\Sigma(U_s \cdot \mu_s)}}{|\Sigma \mu_s|} \quad (5)$$

where $U_s$ and $\mu_s$ are the percentage and quantity (daily mean emissions) of the uncertainty of the sector, respectively.

**4.4 Correlation Analysis**

1) Data sources

The daily deaths of COVID-19 by countries is collected from Worldometers (https://www.worldometers.info/coronavirus/). The stringency index is collected from the OxCGRT project (https://github.com/OxCGRT/covid-policy-tracker), ranging between 0 and 100 to indicate the level of government response (mainly closure measures and containments). The mobility trend of place of residences is collected from Google Mobility Report (https://www.google.com/covid19/mobility/), which shows the relative changes of duration people spent at residential places.

2) Pearson Correlation Coefficient

We calculate the Pearson Correlation Coefficients to measure the linear correlation of every two sets of data. The coefficient is calculated as follow:

$$r = \frac{\Sigma (x - \underline{x})(y - \underline{y})}{\sqrt{\Sigma (x - \underline{x})^2} \sqrt{\Sigma (y - \underline{y})^2}}$$

**4.5 Temperature Correction**

Temperature is one of the most important factors that are affecting the daily emission changes. Temperature would affect the demand for electricity and gas usage and ultimately affects the fuel consumption. The mechanism between temperature and emissions is complicated and unclear, thus we will not make a specific analysis here. However, emission change is still related to the change of temperature (SI Fig3). Therefore, we use polynomial functions (quadratic function) to fit the relationship between temperature and emissions, and calculate the impact of temperature difference on emissions.

First, we fit the relationship between temperature and emissions by using a polynomial model for each country, based on 2019 data only to exclude the anomalies during COVID-19 in 2020 (SI Fig3). In most countries, the model shows good performance ($R2>0.5$). Thus, the emissions difference $\Delta E_d$ caused by the temperature difference should be:

$$\Delta E_d = f(T_{d,2020}) - f(T_{d,2019})$$

Then, the emission after removing the influence of temperature difference should be (SI Fig 4):

$$E'_d = E_d + \Delta E_d$$

The contribution of temperature difference to the daily emissions in day $d$ would be:

$$C = \Delta E_d / E'_d$$

The monthly contribution of temperature difference to the monthly emissions are listed in SI Table 3.

## 5 Data availability

The dataset organized in an Excel file is available at https://doi.org/10.6084/m9.figshare.13685839 (Deng et al., 2021). Country-specific and sector-specific emission data are also available from the Carbon Monitor (https://carbonmonitor.org and http://carbonmonitor.org.cn).

The file CarbonMonitor_DailyCO2_2020.xlsx includes the following:

- Summary table of daily CO2 emissions in 2020 and 2019 of China, India, USA, EU27&UK, Russia, Japan, Brazil, ROW (rest of world), World, UK, France, Germany, Italy and Spain;
- Summary table of daily CO2 emissions in 2020 and 2019 from sectors of power, industry, ground transportation, aviation, residential, and international shipping;
- Country-specific daily CO2 emissions from the power sector in 2020 and 2019;
- Country-specific daily CO2 emissions from the industry sector in 2020 and 2019;
- Country-specific daily CO2 emissions from the ground transportation sector in 2020 and 2019;
- Country-specific daily CO2 emissions from the aviation sector (including domestic and international aviation) in 2020 and 2019;
- Country-specific daily CO2 emissions from the residential sector in 2020 and 2019.

**Author contributions**

ZL, ZD, PC, and SD designed the paper. ZD, JT, BZ coordinated the data processing, ZL, ZD, PC, JT, BZ, SD, RA, OC, FB, PC, XD, PF, PG, RG, CH, RJ, CK, DK, PK, CLQ, CM, GJM, GP, KT, YW, BZ, HZ, HJS contributed to paper writing and revising.

**Acknowledgements**

Authors acknowledge comments by reviewers that helped to improve the paper.

GP and RA received funding from the European Union's Horizon 2020 research and innovation programme under grant agreements No. 776810 (VERIFY), No. 820846 (PARIS REINFORCE), and No. 958927 (CoCO2).

**Data Availability Statement**

All data generated or analyzed during this study are included in this article (and its Data descriptor paper and supplementary information files).

**Code Availability Statement**

The code generated during and/or analyzed during the current study are available from the corresponding author on reasonable request.

**Declaration**

Authors declare no competing interests.

**Supplementary Information**

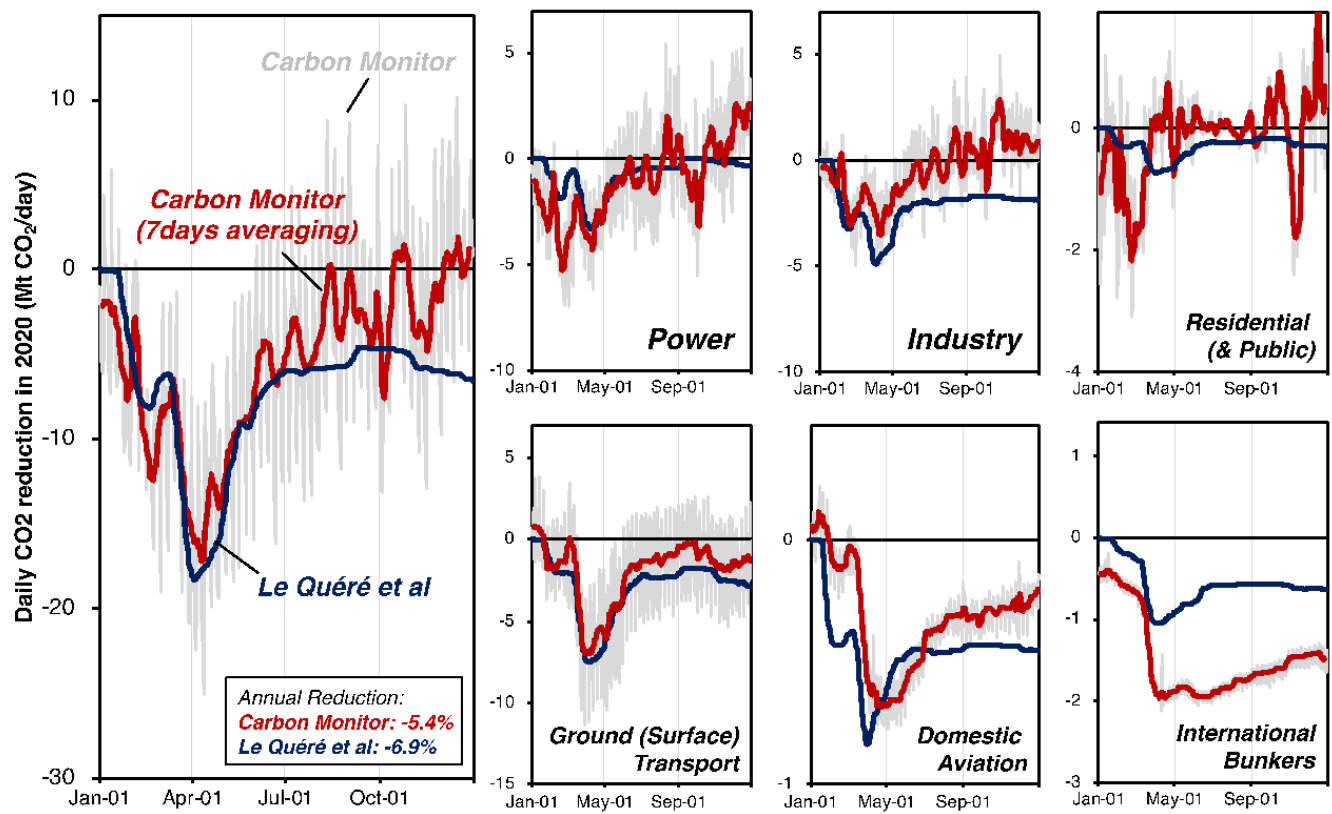

**SI Figure 1 | Comparison of estimates from this study (Carbon Monitor) and Le Quéré et al. (2020) study.** The grey lines denote the daily CO2 reduction in 2020 compared to the same day in 2019 from this study and the read lines denote the 7-day running mean values. The dark blue lines denote the estimation of daily CO2 reduction (7-day running mean) from Le Quéré et al. (2020).

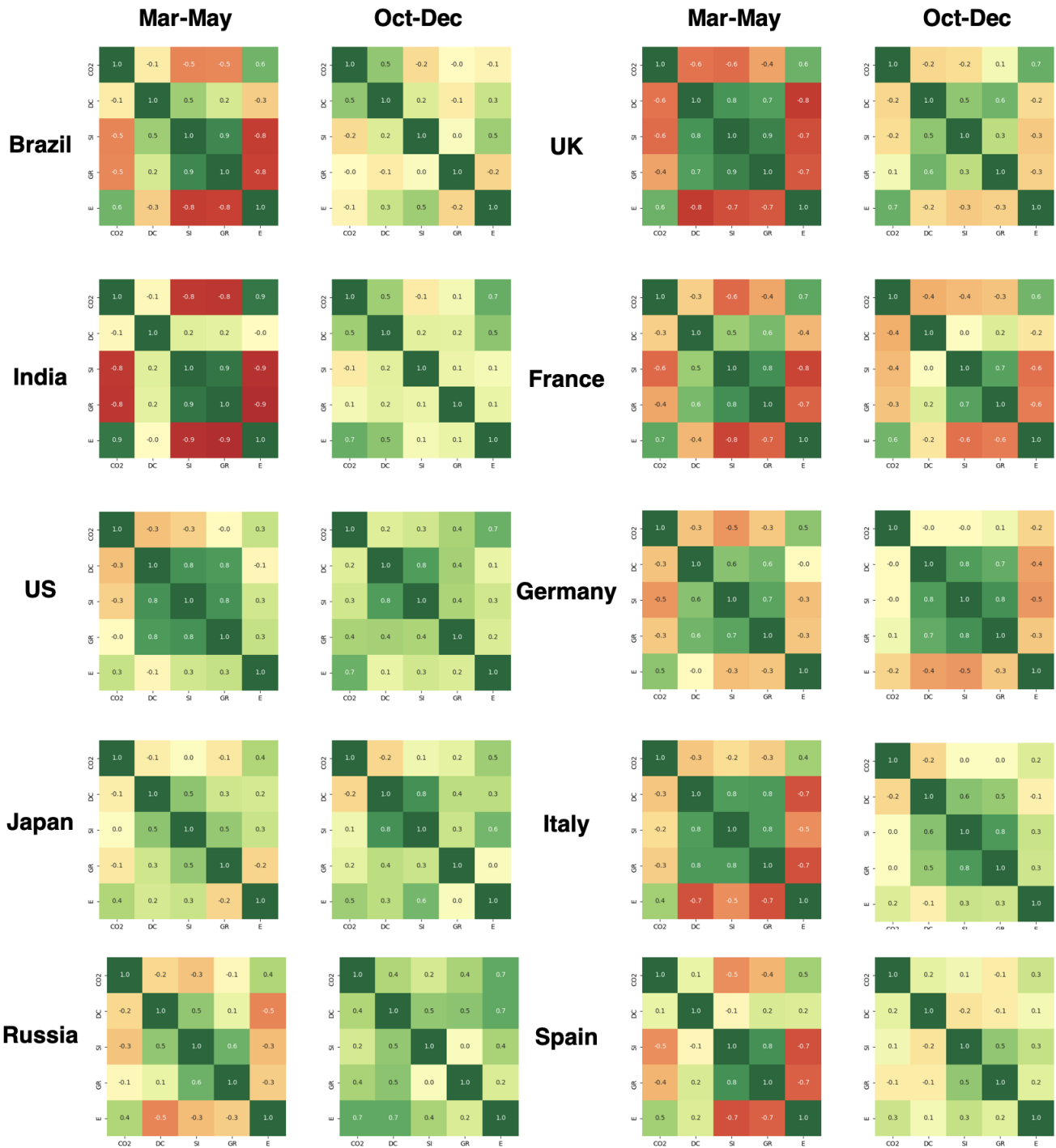

**SI Figure 2 | Correlation matrixes of five indicators during March 1st ~ May 31st, 2020 and October 1st ~ December 31st, 2020.** $CO2$ denotes the daily CO2 changes in 2020 compared to the same day in 2019, $DC$ denotes the daily new cases of COVID-19, $SI$ denotes the

stringency index of government responses to COVID-19, $GR$ denotes the duration spent at places of residence, and $E$ denotes the daily changes of power demand in 2020 compared to the same day in 2019.

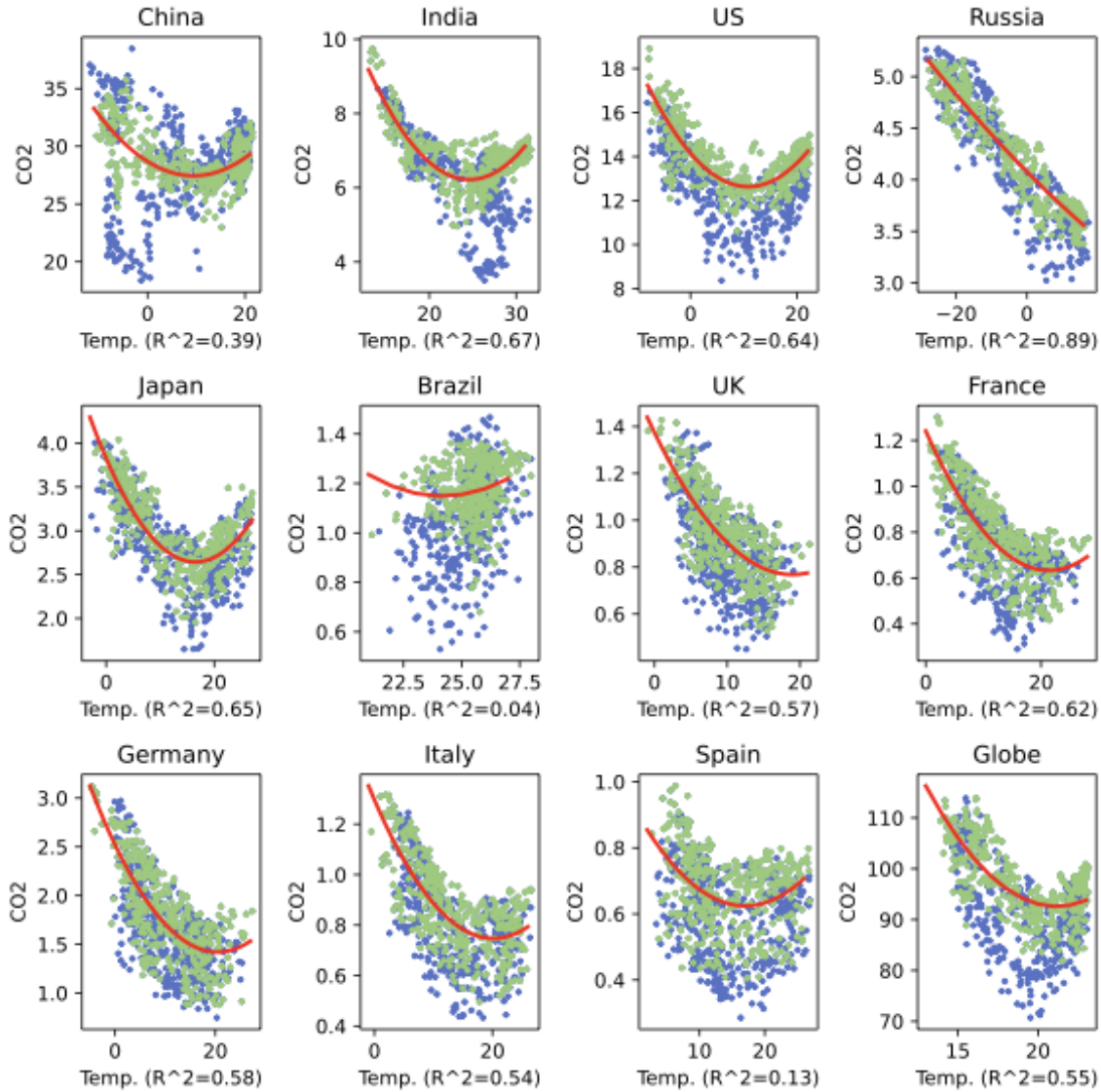

SI Fig 3 | Relationship between daily CO2 emissions and daily temperature in 2019 (green dots) and 2020 (blue dots). The red line represents the fitted polynomial function curve based on only 2019 data.

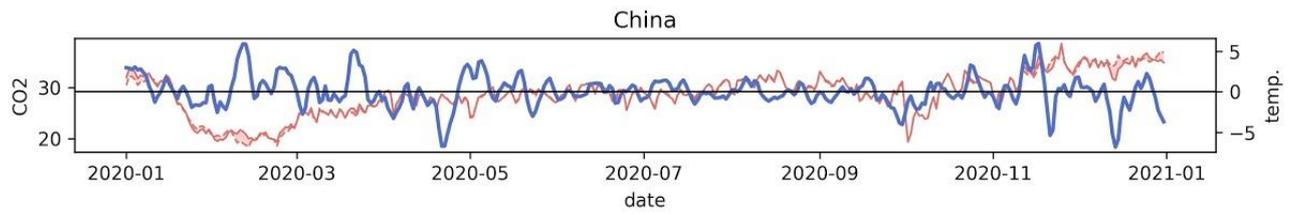
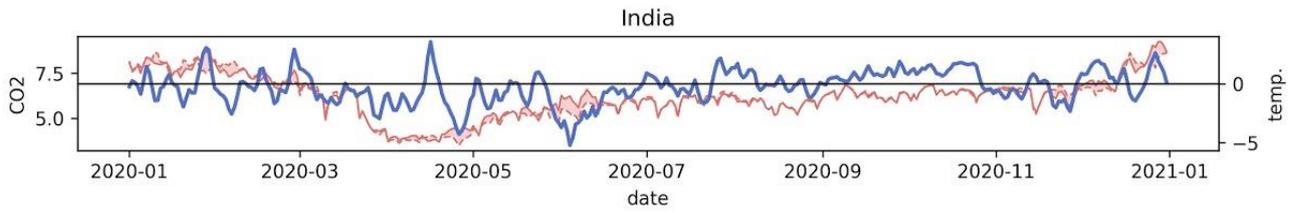
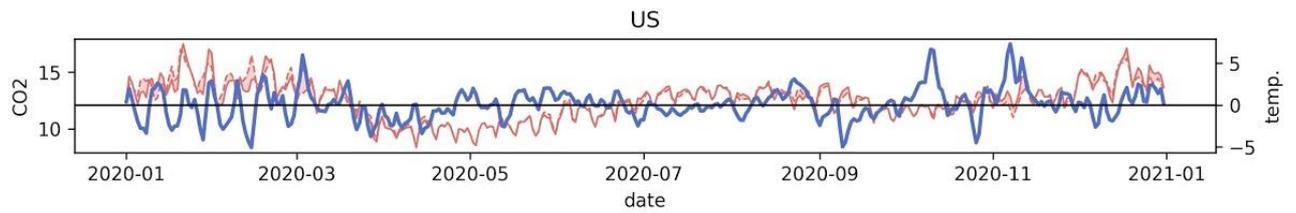
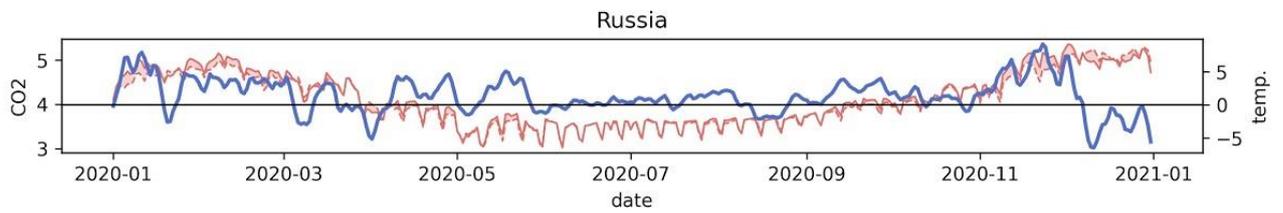
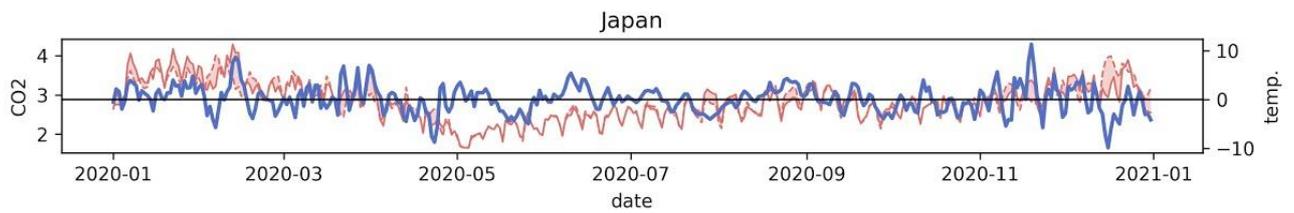
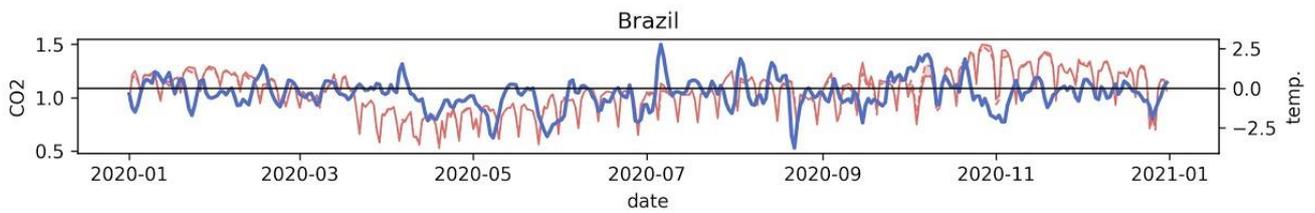
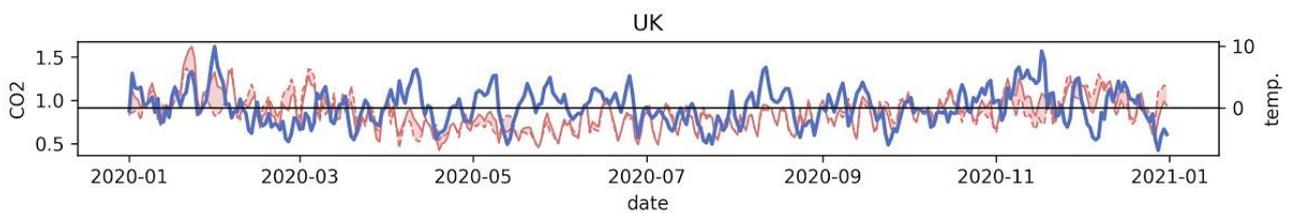
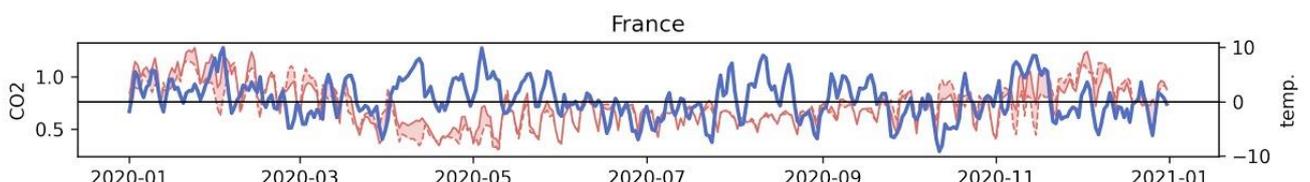

SI Figure 4 | Daily CO2 emissions in 2020 before (red solid line) and after (red dash lines) removing the influence of temperature. The blue lines represent the daily difference of temperature in 2020 compared to the same day in 2019.

**SI Table 1 | Sources of activity data from different sectors**

| Sector | Region | Data type | Source |
|---|---|---|---|
| Power Sector | China | thermal production / coal consumption | China National Power Grid (thermal production) National Bureau of Statistics (thermal production) WIND (coal consumption in Zhedian Company) |
| | India | thermal production | Power System Operation Corporation Limited (https://posoco.in/reports/daily-reports/) |
| | USA | thermal production | Energy Information Administration's (EIA) Hourly Electric Grid Monitor (https://www.eia.gov/beta/electricity/gridmonitor/) |
| | Europe | thermal production | ENTSO-E Transparent platform (https://transparency.entsoe.eu/dashboard/show) |
| | Russia | total Production | United Power System of Russia (http://www.so-ups.ru/index.php) |
| | Japan | thermal production / total demand | OCCTO (Organization for Cross-regional Coordination of Transmission Operators, https://www.occto.or.jp/) Hokkaido Electric Power Network (http://www.rikuden.co.jp/nw/denki-yoho/index.html), Tohoku Electric Power Network (https://setsuden.nw.tohoku-epco.co.jp/download.html), Tokyo Electric Power Company (https://www.tepco.co.jp/forecast/html/download-j.html), Chubu Electric Power Grid (https://powergrid.chuden.co.jp/denkiyoho/), Hokuriku Electric Power Company (http://denkiyoho.hepco.co.jp/area_forecast.html), Kansai Electric Power (https://www.kansai-td.co.jp/denkiyoho/download/index.html), Chugoku Electric Power Transmission & Distribution (https://www.energia.co.jp/nw/jukyuu/index.html), Shikoku Electric Power Company (https://www.yonden.co.jp/nw/denkiyoho/download.html), Kyushu Electric Power Transmission & Distribution (https://www.kyuden.co.jp/td_power_usages/history202010.html), and Okinawa Electric Power Company(http://www.okiden.co.jp/denki2/dl/) |
| | Brazil | Thermal production | Operator of the National Electricity System (http://www.ons.org.br/Paginas/) |
| Industry Sector | China | Industrial product outputs | National Bureau of Statistics (http://www.stats.gov.cn/english/) Trading Economics (https://tradingeconomics.com) |
| | India | Industrial Production Index | Ministry of Statistics and Programme Implementation (http://www.mospi.nic.in) Trading Economics (https://tradingeconomics.com) |
| | USA | Industrial Production Index | Federal Reserve Board (https://www.federalreserve.gov) Trading Economics (https://tradingeconomics.com) |
| | EU | Industrial Production Index | Eurostat (https://ec.europa.eu/eurostat/home) Trading Economics (https://tradingeconomics.com) |
| | Russia | Industrial Production Index | Federal State Statistics Service (https://eng.gks.ru) Trading Economics (https://tradingeconomics.com) |
| | Japan | Industrial Production Index | Ministry of Economy, Trade and Industry Trading Economics (https://tradingeconomics.com) |
| | Brazil | Industrial Production Index | Brazilian Institute of Geography and Statistics (https://www.ibge.gov.br/en/institutional/the-ibge.htm) |

| | | | Trading Economics (https://tradingeconomics.com) |
|---|---|---|---|
| Ground Transportation | 416 Cities | TomTom Congestion Level | TomTom Traffic Index (https://www.tomtom.com/en_gb/traffic-index/) |
| Aviation | World | Flight distance | FlightRadar24 (https://www.flightradar24.com/) |
| Residential | World | population-weighted heating degree days | ERA5 reanalysis of 2-meters air temperature(Copernicus Climate Change Service (C3S), 2019) |
| International Shipping | World | Ship traffic volume | Kinsey, A. Coronavirus Intensifies Global Shipping Risks, https://www.maritime-executive.com/editorials/coronavirus-intensifies-global-shipping-risks (2020). |

**SI Table 2 | Data sources of the U.S. Carbon Monitor Project**

| Sector | Data type | Source |
|---|---|---|
| State-level monthly energy consumption data | | |
| Power sector | fuel-specific consumption data | EIA Electricity (https://www.eia.gov/electricity/data/state/) |
| Industry sector | natural gas consumption | EIA Natural Gas (https://www.eia.gov/naturalgas/data.php) |
| Ground transportation | prime supplier sales volumes of motor gasoline and diesel | EIA Petroleum & Other Liquids (https://www.eia.gov/petroleum/data.php) |
| Aviation | prime supplier sales volumes of kerosene-type jet fuel | EIA Petroleum & Other Liquids (https://www.eia.gov/petroleum/data.php) |
| Residential and commercial | natural gas consumption | EIA Natural Gas (https://www.eia.gov/naturalgas/data.php) |
| State-level daily indicators | | |
| Power sector | thermal generation produced by coal, petroleum and gas | EIA Hourly Electric Grid Monitor (https://www.eia.gov/beta/electricity/gridmonitor/) |
| Industry sector | pipeline deliveries to industrial end users | Genscape Natural Gas (https://www.genscape.com) |
| Ground transportation | distance traveled | Trips by Distance data from Bureau of Transportation Statistics (https://data.bts.gov/Research-and-Statistics/Trips-by-Distance/w96p-f2qv) |
| Aviation | kilometers flown of flights | FlightRadar24 database (https://www.flightradar24.com) |
| Residential and commercial | population-weighted heating degree days | ERA5 reanalysis of 2-meters air temperature (https://cds.climate.copernicus.eu/) |

SI Table 3 | Contribution of (the difference of) temperature in 2020 compared to the same day of 2019 to the daily CO2 emissions. The negative values indicate that relative warmer temperature in this month led to a drop of monthly CO2 emissions; in contrast, the positive values indicate that relative colder temperature in this month contributed a proportion of increase of monthly CO2 emissions.

|  | China | India | US | Russia | Japan | Brazil | UK | France | Germany | Italy | Spain | Globe |
|---|---|---|---|---|---|---|---|---|---|---|---|---|
| Jan | -0.9% | -0.6% | 0.8% | -3.1% | -5.3% | -0.2% | -8.8% | -10.6% | -11.8% | -8.5% | -2.0% | -2.1% |
| Feb | -2.2% | 1.2% | 1.5% | -2.7% | -1.1% | -0.4% | 4.7% | -5.1% | -4.9% | -4.6% | -3.9% | -1.5% |
| Mar | -0.1% | 0.3% | -1.1% | -0.1% | -2.4% | -0.5% | 5.0% | 4.2% | 6.9% | 5.3% | 2.1% | -0.1% |
| Apr | 0.2% | -5.0% | 1.4% | -1.5% | 2.0% | -0.9% | -2.8% | -13.5% | -3.3% | -2.2% | -3.5% | 0.0% |
| May | 0.2% | -3.3% | 0.5% | -1.3% | -0.5% | -0.1% | -3.7% | -8.8% | -2.9% | -8.4% | 0.7% | -0.1% |
| Jun | 0.0% | -6.0% | 0.2% | -0.1% | 1.1% | 0.1% | -1.2% | 0.6% | 2.1% | -0.5% | -1.5% | -0.3% |
| Jul | 0.0% | 0.0% | -0.8% | -0.7% | -2.1% | -0.4% | 3.1% | -0.2% | -0.7% | -0.8% | 1.0% | -0.2% |
| Aug | 0.0% | 0.2% | 1.2% | 0.1% | 2.4% | -0.1% | 0.3% | 1.0% | 0.8% | 0.2% | 0.2% | 0.0% |
| Sep | -0.3% | 0.7% | -1.3% | -1.8% | 0.0% | -1.6% | 1.1% | 0.8% | -1.0% | 1.0% | 0.7% | 0.1% |
| Oct | -0.2% | 0.1% | 0.1% | -0.8% | 0.5% | 1.4% | -1.8% | 6.5% | -0.3% | 4.9% | 1.2% | 0.0% |
| Nov | -0.4% | 2.2% | -2.6% | -3.9% | -2.2% | -1.0% | -10.1% | -7.6% | -3.3% | -0.1% | -4.4% | 0.2% |
| Dec | 1.2% | -3.3% | -1.0% | 1.3% | 4.4% | -0.6% | 4.0% | 3.6% | 2.9% | 2.7% | 4.0% | 1.1% |
| Mar-May | 0.1% | -2.3% | 0.2% | -0.9% | -0.5% | -0.5% | 0.0% | -5.0% | 0.7% | -1.4% | 0.0% | -0.1% |
| Oct-Dec | 0.3% | -0.6% | -1.2% | -1.1% | 1.0% | 0.0% | -2.7% | 0.6% | -0.3% | 2.4% | 0.2% | 0.5% |

**SI Section 1 Comparison of Carbon Monitor emissions with other estimates based on energy data**

To date, not many emissions reports have been made public by the environmental / inventory agencies of the world. We collected available data to assess the performance of the baseline of the Carbon Monitor (year 2019) ex post.

**Department for Business, Energy & Industrial Strategy, UK Government**

The Department for Business, Energy & Industrial Strategy (DBIS) of the UK provides an emission report by quarter up to 2019[1]. Although the breakdown of emissions by sector is not the same as the Carbon Monitor, they are both based on UNFCCC standards and as such, the DBIS data is a good benchmark for the Carbon Monitor.

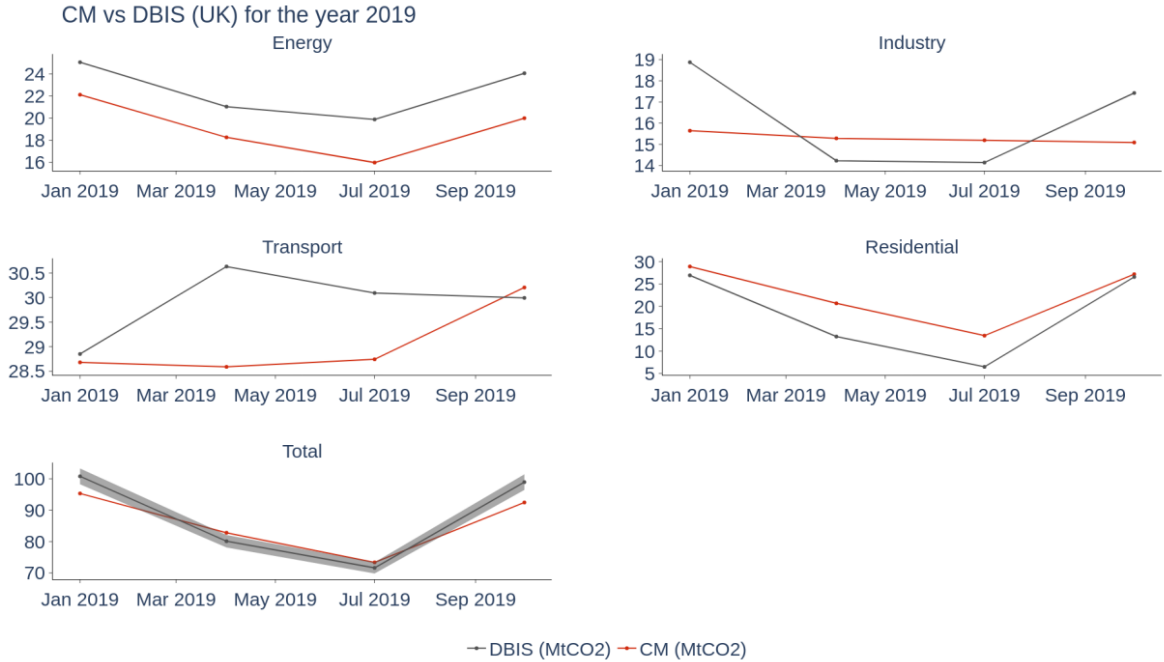

SI Figure 5: Carbon Monitor sectoral emissions compared to DBIS data for 2019. Sectors have been summed when appropriate. Shadowed area corresponds to 95% confidence interval

The carbon dioxide emission total of the Carbon Monitor for each of the quarters of 2019 is in line with the DBIS total, with a maximum 7% deviation. In all sectors except the Transport and Industry sectors, seasonality is comparable. The discrepancy for the Industry sector (called "Business" in DBIS) with higher emissions in quarter 1 and quarter 4 in DBIS can be explained by the fact that DBIS takes heating of businesses and industry buildings into account in this category, while the Carbon Monitor does not (see 2). The data in SI Figure 6 show that the CM allocates 15% less emissions in the first and last quarter and 7% more to the second and third quarters compared to DBIS, although the annual DBIS and CM emissions for their respective definitions of the 'industrial' sector differ only by 5.4%.

---

[1] https://www.gov.uk/government/collections/provisional-uk-greenhouse-gas-emissions-national-statistics

The largest CM discrepancy to DBIS for the seasonality seems to be on the Transport side: DBIS takes domestic ship transport into account while CM does not. More importantly, CM is based on an emission spatial weighted average (from EDGARv4.3.1) of temporal daily emissions from 25 cities with TomTom data, and thus ignores the contribution of other cities. This misfit of seasonality will be investigated in further work involving more precise geolocation data covering all urban areas and other roads transport.

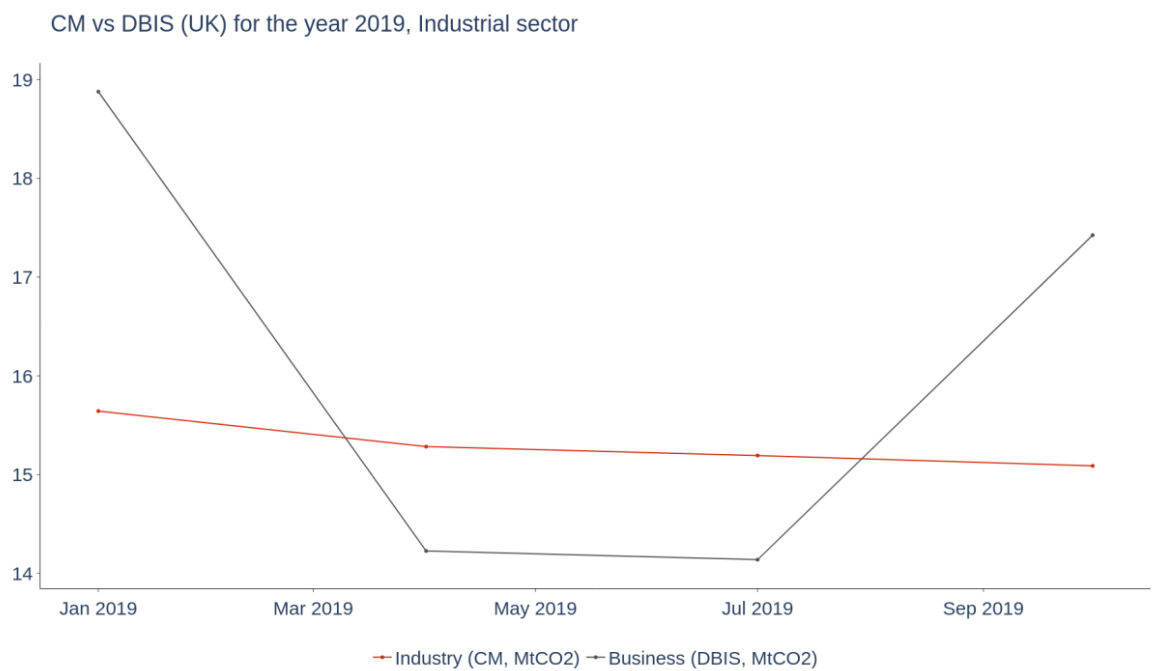

SI Figure 6 Carbon Monitor "Industry" emissions compared to DBIS "Business" emission

**India's monthly fossil CO2 emissions[2]**

Based on early data on fossil fuel use from the Indian government, R. Andrew [2] provided estimations of India's emissions in 2019 and 2020, available at [2]. The 2020 estimation of CM is very much in line with RA, given an annual mean relative difference of 8.7%. The 2019 estimation of CM is however systematically lower than RA from February to June (SI Figure 7). The annual mean relative difference in 2019 is 7.9%. This negative bias of CM in the first half of 2019 can be explained by the fact that this approach uses a constant emission factor for the power production, whereas, in reality, the fuel mix is evolving on the scale of days to months. The share of coal in the Indian electricity mix (SI Figure 9) indeed rose unusually from February to June 2019, hence a greater emission factor for electricity during that period. Because CM emissions have a small bias of the same sign during both years, the difference between the two years is more consistent between CM and RA (SI Figure 8).

---

[2] https://essd.copernicus.org/preprints/essd-2020-152/essd-2020-152.pdf

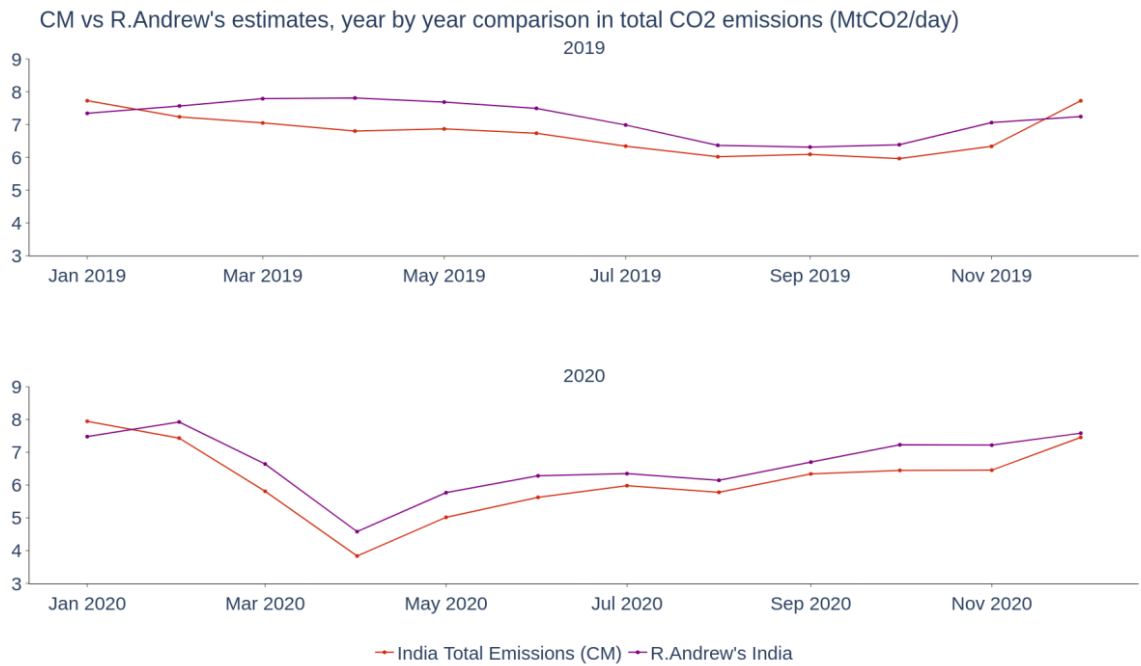

SI Figure 7: Carbon Monitor compared to estimations provided by RA

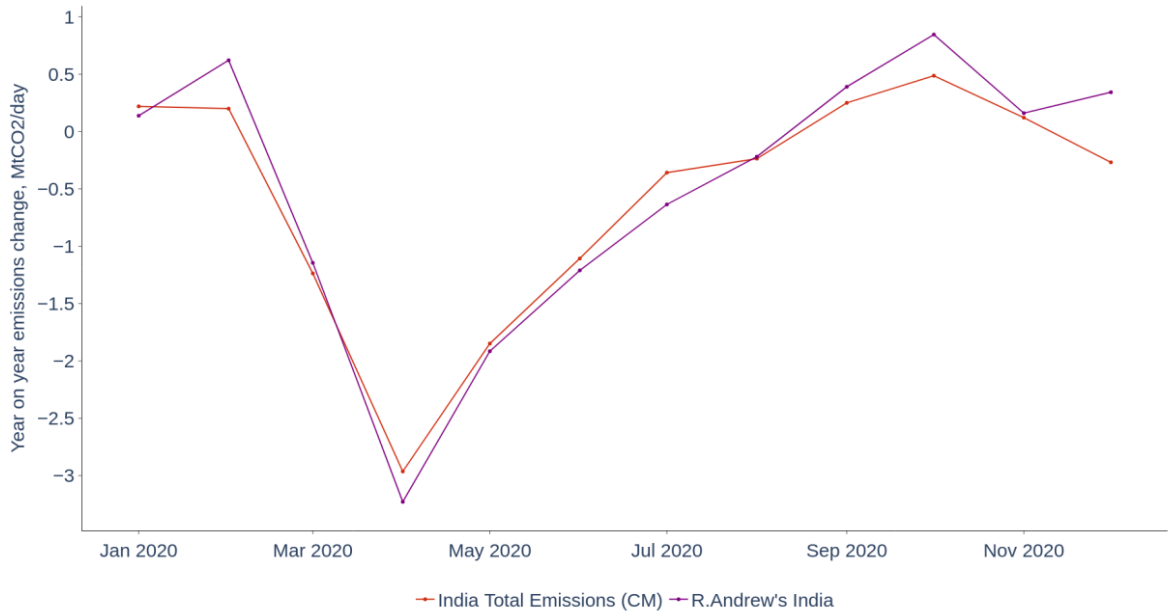

SI Figure 8. Carbon Monitor compared to estimations provided by RA, year-on-year changes 2020 minus 2019.

The estimation of CM is close to RA, given an annual mean relative difference of 8.7% in 2020 and 7.9% in 2019. The small systematic errors between the two estimations are likely due to differences in methodology. While RA accounts for varying fuel mix, CM uses a constant average emission factor. Indeed, India's fuel mix is changing on the scale of months and the higher emissions in the first half of 2019 in Fig S3 can be explained by the higher share of coal in SI Figure 9.

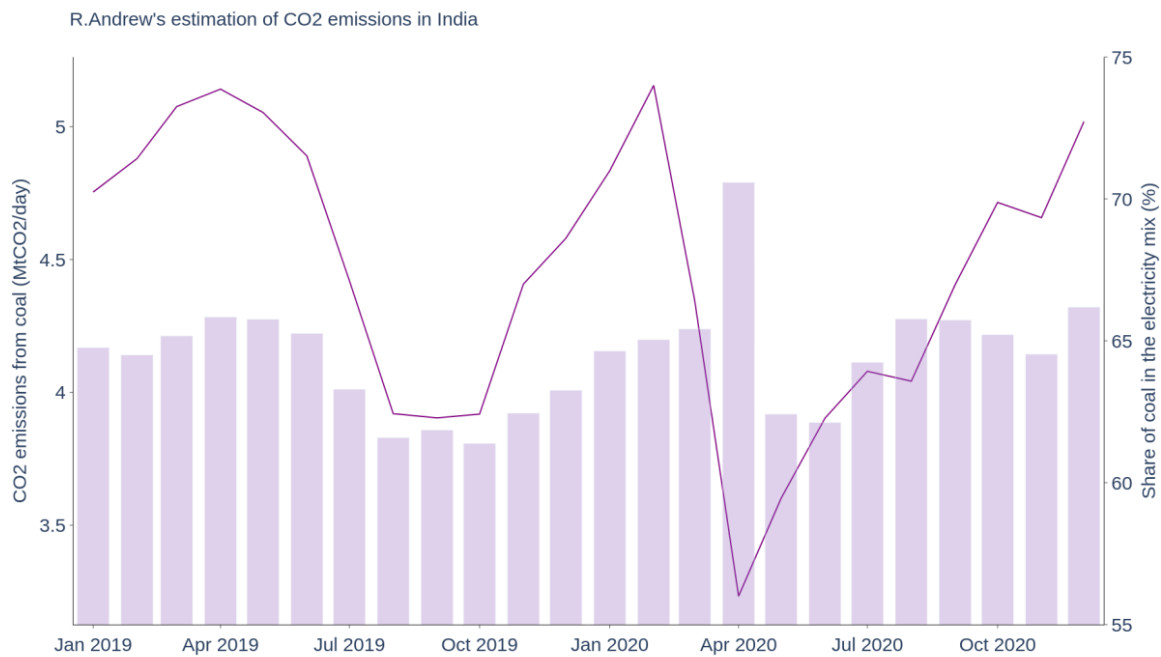

SI Figure 9. Share of coal in the electricity mix of India

**Energy Information Agency, USA**

The EIA regularly publishes CO2 emissions estimates for the US in its Monthly Energy Review[3] publication, and associated data. Recently the Carbon Monitor was enhanced[4] by using more of the EIA data as a baseline in 2017 and 2018, especially the sectoral breakdown of emissions (Peng et al. in preparation). As a result, the latest version of the Carbon Monitor matches EIA data for 2019 and 2020 used in this study much more closely than the older version of Liu et al., with a mean relative bias going from 34.8% MAE in 2019-2020 changes (older version) to 12.9% (latest version). Computations of absolute levels of emissions are also greatly improved as well. The table below presents these improvements.

| Year | MAE (%) | RMSE (%) |
|---|---|---|
| 2019, older method | 5.1 | 6.1 |

---

[3] https://www.eia.gov/totalenergy/data/monthly/#environment
[4] "Carbon Monitor, a near-real time daily dataset of global CO2 emissions from fossil fuel and cement production", revised 20-02-2020

| 2019, latest method | 1.7 | 0.74 |
| 2020, older method | 6.7 | 7.1 |
| 2020, latest method | 1.78 | 0.83 |

Table S3. Mean absolute error (MAE) and Root Mean Square Error (RMSE) between EIA emissions and CM older (Liu et al. 2020) and latest versions ( this study )

SI Figure 10: Latest vs older method estimates for Carbon Monitor US

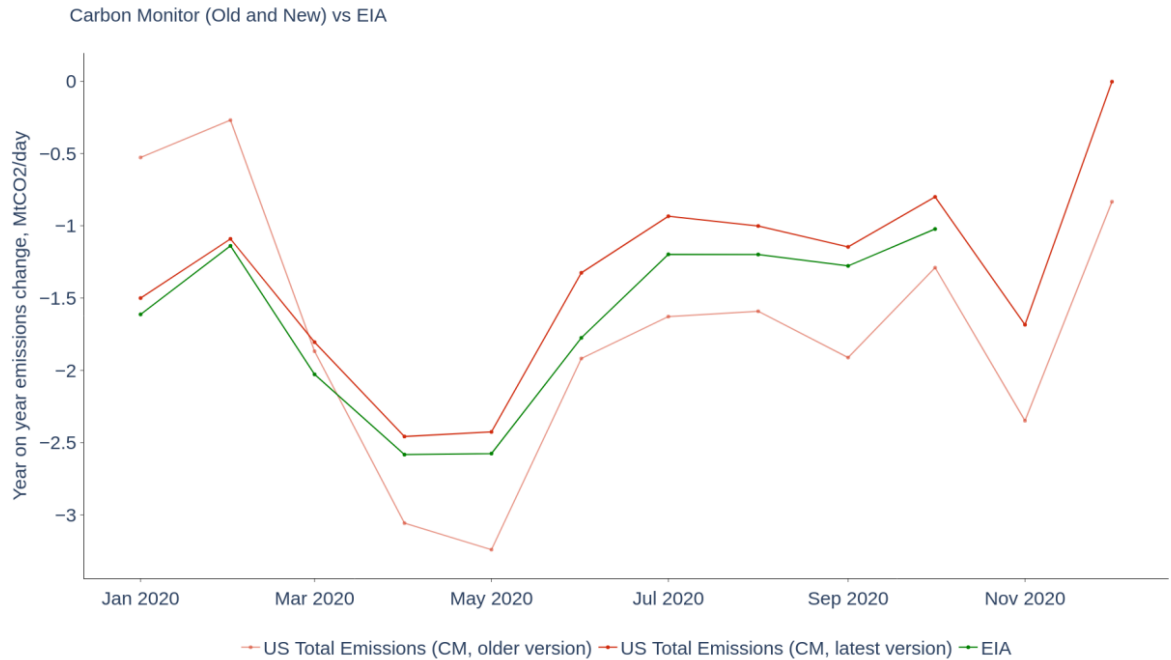

SI Figure 11 CO2 emissions levels in 2019 and 2020

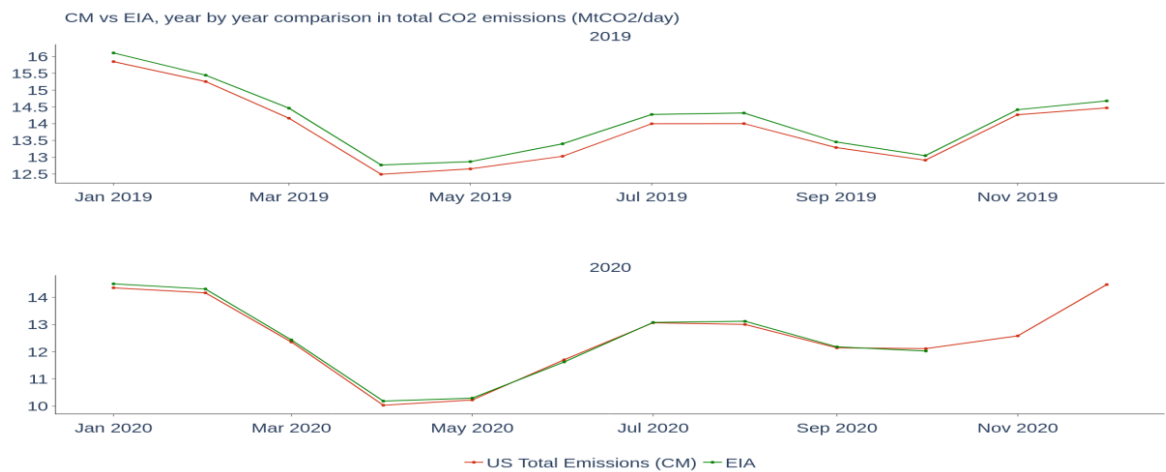